\documentclass[12pt,a4paper]{article}
\usepackage{graphicx} % Required for inserting images
\usepackage{physics}
\usepackage{amsmath}
\usepackage{amssymb} % for \mathbb{holloword}
\usepackage{amsfonts}
\usepackage{hyperref} % For hyperlinks
\usepackage{subcaption} % For Better subfigure 
\usepackage{authblk}
\usepackage{cite} 
\usepackage{authblk}

\begin{document}

% Custom Title Block
\begin{center}
    {\LARGE \textbf{Bloch Sphere of the Qutrit System}} \\[1.5em]
{\large Surajit Sen\textsuperscript{\dag} and Tushar Kanti Dey \textsuperscript{\ddag}} \\[0.5em]
    Centre of Advanced Studies and Innovation Lab, 18/27 Kali Mohan Road, Tarapur, Silchar 788003, INDIA \\[1em]
% Corresponding Author Email
    \textsuperscript{\dag}Corresponding Author: \texttt{ssen55@yahoo.com}\\[1.5em]
% Emails
    \textsuperscript{\ddag}Email: \texttt{tkdey54@gmail.com} \\[0.5em]
%Date  
%    {\small \today} % Date

\end{center}

\vspace{2cm}

\begin{abstract}
We present a novel method to study the Bloch space of the qutrit system by examining the Bloch trajectories in it. Since such system is inherently a three-level quantum system, therefore we use the SU(3) group as the basis group to obtain the Bloch vectors of different configurations of it. The norm of the Bloch space is evaluated from the geometric consideration and also from the dynamics of the Bloch vectors and both results are found to be identical. The analysis of the dynamical evolution of the Bloch vectors reveals an additional feature that, under resonant conditions, the Bloch sphere $\mathbb{S}^{7}$ splits into two parts, a four-sphere $\mathbb{S}^{4}$ and a two-sphere $\mathbb{S}^{2}$. The Bloch trajectories of the two sectors across different configurations exhibit a range of simple to complex curves, highlighting the non-trivial structure of the Bloch space of the qutrit system.
\end{abstract}


\vfill
%=======================================

\newpage 
\section{Introduction}
\label{sec1intro}
\par 
The Bloch sphere representation is possibly the most elegant way of visualizing the state-space structure of the two-level quantum system \cite{MilonniEberly2010}. Such system, which is commonly known as qubit, is geometrically represented by a unit two-sphere $\mathbb{S}^2$ embedded in real three-dimensional space $\mathbb{R}^{(3)}$ with poles $\ket{0}$ and $\ket{1}$ as its two principal cardinal states \cite{MilonniEberly2010,Nielsen2010}. All other states on the surface of the Bloch sphere can be built-up by the superposition of these cardinal states. Quantitatively, the Bloch sphere representation is characterized by the complex-valued density matrix operator \cite{Benenti2004,Jaeger2006},
\begin{align}
    \rho_B = \frac{1}{2} \big(\sigma_0 + \mathbf{n}_B \cdot \boldsymbol{\sigma}\big), \quad \mathbf{n}_B \in \mathbb{R}^3
    \label{eq1}
\end{align}
where $\sigma_0$ is the identity matrix and $\sigma_i$ ($i=1,2,3$) represents the Pauli matrices. The Bloch vector of the qubit system  $\mathbf{n}_B$ appearing in Eq.\eqref{eq1} is defined as, 
\begin{align}
\begin{aligned}
    \mathbf{n}_B =& \text{Tr}[\rho_B \, \boldsymbol{\sigma}], \quad {n_1}^2 + {n_2}^2 + {n_3}^2 = |\mathbf{r}_n|^2, \quad \mathbf{r}_n \in \mathbb{R}^3
    \label{eq2}
\end{aligned}
\end{align}
where the density matrix $\rho_B$ satisfies the properties: $\rho_B = \rho_B^{\dagger}$, $\text{Tr}[\rho_B]=1$, $\rho_B^2 \geq 0$ \cite{Benenti2004,Goyal2016}. The properties of the Bloch vectors as well as the density matrix are consistent with the definition of \textit{Purity} of the two-level system \cite{Benenti2004,Jaeger2006},
\begin{align}
\begin{aligned}
    \text{Tr}[\rho_B^2] = &\frac{1}{2} \big(1 + |\mathbf{r}_n|^2\big),  \quad 0 \leq |\mathbf{r}_{n}|^2 \leq 1, \\ 
    &\frac{1}{2} \leq \text{Tr}[\rho_B^2] \leq 1.
    \label{eq3}
\end{aligned}
\end{align}
%which ranges from $\frac{1}{2} \leq \text{Tr}[\rho_B^2] \leq 1$. 
In the realm of quantum information theory, the Bloch sphere plays very crucial role to elucidate different facets of the qubits such as, quantum gates and their operations,  
%using Pauli matrices and their variants \cite{Nielsen2010}, 
addressing measurement-related issues \cite{Nielsen2010}, studying the $N$-level systems known as qudits \cite{Kossakowski2005,Stanislaw2006,Bertlmann2008,Arvind2003,Kimura2008} and, most importantly, the entanglement scenario of the system with two or more qubits  \cite{Jakobczyk2001,Avron2007,Morelli2023,Wie2014,Bley2024}.
\par 
On the other hand, a qutrit system, the immediate generalization of a qubit, is inherently a three-level quantum system defined in Hilbert space $\mathbb{H}^{(3)}$. Since it is knitted with the foundational issues of the quantum mechanics in high-dimension, therefore recently considerable attention has been paid to visualize its Bloch sphere using different theoretical frameworks. These studies include, quaternion-inspired models \cite{Whartonkoch2015}, Heisenberg-Weyl methods \cite{Ali2016,Sharma2024} and other approaches  \cite{Goyal2016, Baumgartner2006,Kurzynski2011,Bengtsson2017,Li2013}. Apart from them, recently there has been growing interest to construct the Bloch sphere for the qutrit system using SU(3) group because of its innate simplicity, rich underlying symmetry and close analogy with the qubit system \cite{Goyal2016, Bertlmann2008, Kurzy2016, Kurzyinski2011,sen2024}. In spite of some progress, these methods become quite complex due to the limited information available about the structure of its Bloch space. One possible way to simplify the complexity of the qutrit system is to evaluate the Bloch vectors from using a dynamical approach. This method involves solving the Schr\"{o}dinger equation for the three-level system and then construct the Bloch vectors from its solution. Implementing this method requires using all possible configurations of the three-level system which are categorized into three types: $\Lambda$, $V$, and $\Xi$ configurations. These models have been extensively studied in recent years and applied to address a wide variety of quantum optical systems \cite{sen2024, Hioe1982, Hioe1985,Yoo1985, Nath2008,Sen2016,Sen2017,Sen2023}. In this paper, we analyze the properties of the Bloch space for each configuration by examining the phase space trajectories of the Bloch vectors, which we will refer to as SU(3) Bloch vectors, and demonstrate how this approach captures the nuances and subtleties of the Bloch sphere that may not be evident through other methods.
\par 
The remaining sections of this paper are structured as follows: In Section II, we consider the standard parametrization scheme of the qutrit system in terms of the spherical polar coordinates from the geometric consideration and obtain the norm of the Bloch space. We then discuss the cardinal states of the system and obtain all possible superposed states from them. Since the qutrit system is essentially a three-level system at the heart, therefore in Section III, we give detailed evaluation of the SU(3) Bloch vectors from the dynamical consideration while taking account of all configurations allowed by the selection rule. In Section IV, we analyze the Bloch trajectories of different configurations from their phase space portraits and discuss their behaviour both at resonance and off-resonance. Finally, we conclude by summarizing the essential content of the paper and discuss the outlook.

%=====================================
\section{Qutrit system} \label{section2}
\subsection{Cardinal states of qutrit system}\label{subsection2.1}
\par 
The wave function of a qutrit system can be written in the standard  basis, 
\begin{align}
\ket{\psi_{T}}=c_1\ket{1}+c_2\ket{2}+c_3\ket{3}, 
\label{eq4}
\end{align}
where the basis states are given by, 
\begin{align}
     \ket{1}=\begin{pmatrix} 1 \\ 0 \\ 0 \end{pmatrix}\;, \quad 
     \ket{2}=\begin{pmatrix} 0 \\ 1 \\ 0 \end{pmatrix},\; \quad 
     \ket{3}=\begin{pmatrix} 0 \\ 0 \\ 1 \end{pmatrix},\;
     \label{eq5}
\end{align}
and the normalized complex amplitudes are parameterized as, 
\begin{subequations}
\begin{align}
c_1&=\cos{\frac{\theta_1}{2}}, \\
c_2&=e^{i\phi_1}\sin{\frac{\theta_1}{2}}\sin{\frac{\theta_2}{2}}, \\
c_3&=e^{i\phi_2}\sin{\frac{\theta_1}{2}}\cos{\frac{\theta_2}{2}},
\label{eq6}
\end{align}
\end{subequations}
with range, $\{\theta_i\} \in [0,\pi]$ and $\{\phi_i\} \in [0,2\pi)$ $(i=1,2)$. The density matrix read off from the qutrit wave function Eq.\eqref{eq4} is given by, 
\begin{align}
    \rho_{T}
    &=\begin{bmatrix} \cos^2\frac{\theta_1}{2} & \frac{1}{2}e^{i\phi_1}\sin{\theta_1}\sin{\frac{\theta_2}{2}} & \frac{1}{2}e^{i\phi_2}\sin{\theta_1}\cos{\frac{\theta_2}{2}} \\ 
     \frac{1}{2}e^{-i\phi_1}\sin{\theta_1}\sin{\frac{\theta_2}{2}} & \sin^2{\frac{\theta_1}{2}}\sin^2{\frac{\theta_2}{2}} & \frac{1}{2}e^{-i(\phi_1-\phi_2)}\sin^2{\frac{\theta_1}{2}}\sin{\theta_2} \\  \frac{1}{2}e^{-i\phi_2}\sin{\theta_1}\cos{\frac{\theta_2}{2}} & \frac{1}{2}e^{i(\phi_1-\phi_2)}\sin^2{\frac{\theta_1}{2}}\sin{\theta_2} &  \sin^2\frac{\theta_1}{2}\cos^2\frac{\theta_2}{2} \end{bmatrix}, 
     \label{eq7}
\end{align}
which satisfies the idempotent condition, $\rho_{T}^2=\rho_{T}$,  giving the pure state condition, $Tr[\rho_{T}^2]=1$. 
%%%%%%%%%%%%%%%%%%%%%%%%%%%%%%%%%
\par 
Before addressing the norm of the Bloch space, we first consider the key cardinal states and their superposed states by the judicious choice of the coordinates. From Eqs.\eqref{eq4} and \eqref{eq5}, the state vectors at some select location of the Bloch sphere are given by, 
\begin{subequations}\label{eq8}
\begin{align}
\ket{\psi_{T}{(\theta_1=0)}}&= \ket{1}, \label{eq8a}\\ 
\ket{\psi_{T}{(\theta_1= \pi,\theta_2= \pi)}}&= e^{ i\phi_1}\ket{2}, \label{eq8b}\\
\ket{\psi_{T}{(\theta_1=\pi,\theta_2=0)}}&= e^{i\phi_2}\ket{3}, \label{eq8c}
\end{align}
\end{subequations}
which may be referred as three principal cardinal states of the SU(3) Bloch space. Proceeding similar way, it is straight forward to obtain the superposition of these cardinal states, 
\begin{subequations} \label{eq9}
\begin{align}
\ket{\psi_{T}{(\theta_1=\pi/2,\theta_2= \pi)}}&=\frac{1}{\sqrt{2}}\big[\ket{1} + e^{i\phi_2}\ket{2}\big], \label{eq9a}\\
\ket{\psi_{T}(\theta_1=\pi,\theta_2=\pi/2)}&=\frac{1}{\sqrt{2}}\big[ e^{-i\phi_1}\ket{2} + e^{-i\phi_2}\ket{3}\big], \label{eq9b}\\
\ket{\psi_{T}{(\theta_1= \pi,\theta_2=\pi/2})}&=\frac{1}{\sqrt{2}}\big[\ket{1} - e^{i\phi_2}\ket{3}\big], \label{eq9c}
\end{align}
\end{subequations}
for any pair of states and 
\begin{align}
\ket{\psi_{T}{(\theta_1=\pi/2,\theta_2= \pi/3)}}=\frac{1}{\sqrt{2}}\big[\ket{1} + \frac{1}{2}e^{i\phi_1}\ket{2} + \sqrt{\frac{3}{4}} e^{i\phi_2}\ket{3}\big],
\label{eq10}
\end{align}
for all three states. The presence of such nontrivial states, unlike in the qubit system \cite{MilonniEberly2010}, is a primary signature of the complex dynamics of the SU(3) Bloch vectors of the qutrit system, which we will discuss by examining their Bloch trajectories. 

%====================================

\subsection{SU(3) Bloch vectors from geometric consideration}\label{subsection2.2}
\par 
The Bloch vectors for the qutrit system in the SU(3) basis is given by \cite{Goyal2016, Sen2012},  
\begin{align}
\mathbf{S}_{T}=Tr[\boldsymbol{\lambda}\rho_{\text{T}}], \quad \mathbf{\text{S}}_{T} \in \mathbb{R}^{(8)}, 
\label{eq11}
\end{align}
where $\rho_{\text{T}}$ be the density matrix of the  qutrit system given by Eq.\eqref{eq7} and $\lambda_i$ ($i=1,2,\dots 8$) be the GellMann matrices which in the qutrit basis are given by (including unit matrix $\lambda_0$) \cite{Greiner2012},
\begin{equation}
\begin{aligned}
\lambda_0&=\ket{1}\bra{1}+\ket{2}\bra{2}+\ket{3}\bra{3}, \qquad 
\lambda_1=\ket{1}\bra{2}+\ket{2}\bra{1} \\
\lambda_2&=i\big[\ket{2}\bra{1}-\ket{1}\bra{2}\big], 
\qquad  
\lambda_3=\ket{1}\bra{1}-\ket{2}\bra{2}, \\
\lambda_4&=\ket{1}\bra{3}+\ket{3}\bra{1}, \quad 
\qquad 
\lambda_5=i\big[\ket{3}\bra{1}-\ket{1}\bra{3}\big], \\ 
\lambda_6&=\ket{2}\bra{3}+\ket{3}\bra{2}, 
\qquad 
\lambda_7=i\big[\ket{3}\bra{2}-\ket{2}\bra{3}\big], \\ 
\lambda_8 & =\frac{1}{\sqrt{3}}\big[\ket{1}\bra{1}+\ket{2}\bra{2}-2\ket{3}\bra{3}\big].
%\label{12}
\end{aligned}
\label{eq12}
\end{equation}
The matrices are normalized as $\lambda_l\lambda_m=\delta_{lm}+d_{lmn}\lambda_n+f_{lmp}\lambda_p$, where $d_{lmn}$ and $f_{lmp}$ ($l,m,n,p=1,2,\dots,8$) are the completely symmetric and completely antisymmetric structure constants of the SU(3) group. It is straight forward to find the SU(3) Bloch vectors in terms of the angular parameters by plugging in Eqs.\eqref{eq7} and \eqref{eq12} into Eq.\eqref{eq11},
\begin{subequations} \label{eq13}
\begin{align}
    {S_{T_1}}&=\sin\theta_1\sin{\frac{\theta_2}{2}}\cos{\phi_1}, \label{eq13a}\\
    {S_{T_2}}&=-\sin\theta_1\sin{\frac{\theta_2}{2}}\sin{\phi_1}, \label{eq13b}\\
    {S_{T_3}}&= \cos^2{\frac{\theta_1}{2}}-\sin^2{\frac{\theta_1}{2}}\sin^2{\frac{\theta_2}{2}} \label{eq13c}\\
    %\frac{1}{4}\big[1+\cos\theta_1(\cos\theta_2-3)+\cos\theta_2\big],\\
    {S_{T_4}}&=\sin\theta_1 \cos{\frac{\theta_2}{2}} \cos\phi_2, \label{eq13d}\\ 
    {S_{T_5}}&=-\sin\theta_1 \cos{\frac{\theta_2}{2}} \sin\phi_2, \label{eq13e}\\
    {S_{T_6}}&=\sin^2{\frac{\theta_1}{2}}\sin\theta_2\cos(\phi_1-\phi_2), \label{eq13f}\\
    {S_{T_7}}&=\sin^2{\frac{\theta_1}{2}}\sin\theta_2\sin(\phi_1-\phi_2),\label{eq13g}\\
    {S_{T_8}}&=\frac{1}{2\sqrt{3}}\big[(1-3\cos\theta_2)+3\cos\theta_1(1+\cos\theta_2)\big]\label{eq13h},
\end{align}
\end{subequations}
which follow the equation of seven-sphere $\mathbb{S}^7$ embedded in the 8-dimensional Bloch space $\mathbb{R}^8$, 
\begin{align}
    {S}^2_{T_1} + {S}^2_{T_2} + {S}^2_{T_3} + {S}^2_{T_4} + {S}^2_{T_5} + {S}^2_{T_6} + {S}^2_{T_7} + {S}^2_{T_8} = \frac{4}{3}. 
     \label{eq14}
\end{align}
The reason for not scaling the Bloch vectors to have a unit sphere will become clear shortly. 
%=======================================

\section{SU(3) Bloch vectors from dynamical consideration}\label{section3}
\subsection{Three-level system revisited}\label{subsec31}
\par 
To obtain the time-dependent SU(3) Bloch vectors, let us briefly recall dynamics of different configurations of the three-level system. The Hamiltonian of a generic three-level  configuration ($A=\Lambda, V, \Xi$) is given by \cite{Nath2008,Sen2012}, 
\begin{align}\label{eq15}
H^A(t)=H_0^A+H_I^A(t). 
\end{align}
For the $\Lambda$ configuration ($A=\Lambda$), the unperturbed and interaction term of the Hamiltonian are given by, 
\begin{subequations}\label{eq16}
\begin{align}
H_{\text{0}}^\Lambda&= \omega_{13} V_3 +\omega_{23} T_3\label{eq16a}\\ 
H_{\text{I}}^\Lambda(t)&=\frac{1}{2}\kappa_{13}
V^+e^{i\Omega_{13}t} + \frac{1}{2}\kappa_{23} 
T^+e^{i\Omega_{23}t} + h.c.,\label{eq16b}
\end{align}
\end{subequations}
where $\omega_{ij} =|\omega_{i}-\omega_{j}|$ be the difference of energy between $\ket{i}$-th and $\ket{j}$-th level, $\kappa_{ij}$ (i,j=1,2,3 with $i\neq j$) be the atom-field coupling parameters and $\Omega_{ij}$ be the frequency of the external field. Similarly we have, 
\begin{subequations} \label{eq17}
\begin{align}
H_0^V&= \omega_{13} V_3 +\omega_{12} U_3 \label{eq17a} \\ 
H_\text{I}^V(t)&=\frac{1}{2}\kappa_{13}
V^+e^{i\Omega_{13}t} + \frac{1}{2}\kappa_{12}
U^+e^{i\Omega_{12}t} + h.c., \label{eq17b}
\end{align}
\end{subequations}
for the $V$ configuration ($A=V$) and 
\begin{subequations}\label{eq18}
\begin{align}
H_\text{0}^\Xi&=\omega_{12} U_3 +\omega_{23} T_3 \label{eq18a}
\\
H_\text{I}^\Xi(t)&=\frac{1}{2}\kappa_{12}
U^+e^{i\Omega_{12}t} + \frac{1}{2}\kappa_{23}
T^+e^{i\Omega_{23}t} + h.c.\label{eq18b}
\end{align}
\end{subequations}
for the $\Xi$ configuration ($A=\Xi$).  In defining above Hamiltonians, we have the following SU(3) shift operators \cite{Greiner2012},
\begin{align}
T_+ &=\ket{1}\bra{2}, \quad T_- =\ket{2}\bra{1} \quad
T_3=\ket{1}\bra{1}-\ket{2}\bra{2} \nonumber \\ %=\lambda_3, 
V_+ &=\ket{1}\bra{3}, \quad V_- =\ket{1}\bra{2} \quad
V_3=\ket{1}\bra{1}-\ket{3}\bra{3}  \\ % =\frac{1}{2}(\sqrt{8}\lambda_8+\lambda_3),\\
U_+ &=\ket{2}\bra{3}, \quad U_- =\ket{3}\bra{2}, \quad
U_3=\ket{2}\bra{2}-\ket{3}\bra{3} \nonumber % =\frac{1}{2}(\sqrt{8}\lambda_8-\lambda_3). \nonumber 
%\label{eq19}
\end{align}
which obey the following algebra, 
\begin{align} \label{eq20}
[T^+,T^-]&=T^3 \quad [T^+,T^3]=-2T^- \quad [T^-,T^3]=2T^+, \nonumber \\
[V^+,V^-]&=V^3 \quad [V^+,V^3]=-2V^- \quad [V^-,V^3]=2V^+, \\ 
[U^+,U^-]&=U^3 \quad [U^+,U^3]=-2U^- \quad [U^-,U^3]=2U^+, \nonumber
%\label{eq20}
\end{align}
and their operation on the basis states are given by, 
\begin{align} \label{eq21}
     T_{+}\ket{1}&=0, \quad T_{-}\ket{1}=\ket{2} \quad T_3\ket{1}=\ket{1}, \nonumber \\
     T_{+}\ket{2}&=\ket{1}, \quad T_{-}\ket{2}=0 \quad T_3\ket{2}=-\ket{2},  \nonumber  \\
     T_{+}\ket{3}&=0, \quad T_{-}\ket{3}=0 \quad T_3\ket{3}=0, 
     \nonumber \\
     V_{+}\ket{1}&=0, \quad V_{-}\ket{1}=\ket{3} \quad V_3\ket{1}=\ket{1}, \nonumber \\
     V_{+}\ket{2}&=0, \quad V_{-}\ket{2}=0 \quad V_3\ket{2}=0,  \\  
     V_{+}\ket{3}&=\ket{1}, \quad V_{-}\ket{3}=0 \quad V_3\ket{3}=-\ket{3}, \nonumber \\
     U_{+}\ket{1}&=0, \quad U_{-}\ket{1}=0 \quad U_3\ket{1}=0,\nonumber \\ 
     U_{+}\ket{2}&=0, \quad U_{-}\ket{2}=\ket{3} \quad U_3\ket{2}=\ket{2}, \nonumber \\ 
     U_{+}\ket{3}&=\ket{2}, \quad U_{-}\ket{3}=0 \quad U_3\ket{3}=-\ket{3}. \nonumber 
%     \label{eq21}
\end{align}
\par 
Using the unitary matrix for the $\Lambda$ configuration \cite{Sen2012}, 
\begin{equation} \label{eq22}
U_{\Lambda}(t)=\exp\bigg[-\frac{i}{3}
\big((2\Delta_{13}-\Delta_{23})V_3t
+(\Delta_{13}-2\Delta_{23})T_3t\big)\bigg], 
\end{equation}
the Hamiltonian Eq.\eqref{eq16} can be expressed in the time-independent rotating frame \cite{Sen2012}, 
\begin{align}
\hat{H}_\text{R}^\Lambda(0)&=-i\dot{U}^\dag_\Lambda(t)U_\Lambda(t)+U^\dag_\Lambda(t)H^{\Lambda}(t)U_\Lambda(t) \nonumber \\
&=\begin{bmatrix}
\frac{1}{3}(\Delta_{13}+\Delta_{23}) & \kappa_{23} & \kappa_{13} \\
\kappa_{23} & \frac{1}{3}(\Delta_{13}-2\Delta_{23}) & 0 \\
\kappa_{13} & 0 & -\frac{1}{3}(2\Delta_{13}-\Delta_{23}) 
\label{eq23}
\end{bmatrix}
\end{align}
where the detuning offset are given by  $\Delta_{13}=2\omega_{13}+\omega_{13}-\Omega_{13}$ and $\Delta_{13}=2\omega_{13}+\omega_{13}-\Omega_{13}$. Finally taking equal-detuning condition, namely,  $\Delta_{13}=\Delta_{23}=\Delta_{}$, and making use of Equations \eqref{eq20} and \eqref{eq21}, the Hamiltonian \eqref{eq23} leads to normalized amplitudes of the $\Lambda$ configuration, 
\begin{subequations}  \label{eq24}
\begin{align}
c_1(t)&=e^{-\frac{1}{6} i \Delta t} \left\{ 
c_{10} \cos\left(\frac{\Omega t}{2}\right) 
- \frac{i}{\Omega} \left(c_{10} \Delta + c_{30} \kappa_{13} + c_{20} \kappa_{23}\right) \sin\left(\frac{\Omega t}{2}\right) 
\right\}\\
%c_1(t)&=e^{-\frac{1}{6} i\Delta t} \Big[\text{c}_{10}\cos{\left( \frac{\Omega t}{2} \right)}
%-\frac{i}{\Omega}(\text{c}_{10} \Delta +\text{c}_{30}
%   \kappa 2\text{c}_{20} \kappa_{23})\sin{\left( \frac{\Omega t}{2} \right)}\Big] \label{eq24a} \\ 
%\end{align}
%
%\begin{align}
c_2(t)&= \frac{e^{-\frac{1}{6}i \Delta t}}{(\kappa_{13}^2 + \kappa_{23}^2) \Omega} \Big[
e^{\frac{1}{2}i \Delta t} \kappa_{13} \left( c_{20} \kappa_{13} - c_{30} \kappa_{23} \right) \Omega \nonumber\\
&\quad + \kappa_{23} \left( c_{30} \kappa_{13} + c_{20} \kappa_{23} \right) \Omega \cos{\left( \frac{\Omega t}{2} \right)} \nonumber\\
&\quad + i \kappa_{23} \left( c_{30} \Delta \kappa_{13} + c_{20} \Delta \kappa_{23} - c_{10} (\kappa_{13}^2 + \kappa_{23}^2) \right) \sin{\left( \frac{\Omega t}{2} \right)} \Big] \label{eq24b}\\
%\end{align}
%
%\begin{align}
c_3(t)&=-\frac{e^{-\frac{1}{6}i\Delta t }}{\kappa_{13}^2 + \kappa_{23}^2} \Big[
e^{\frac{1}{2}{i \Delta t}} \kappa_{ 23} \left( -c_{20} \kappa_{13} + c_{30} \kappa_{23} \right) \nonumber\\
&\quad + \kappa_{13} \left( c_{30} \kappa_{13} + c_{20} \kappa_{23} \right) \cos^2{\left( \frac{\Omega t}{2} \right)} \nonumber\\
&\quad + i \kappa_{13} \left( c_{30} \Delta \kappa_{13} + c_{20} \Delta \kappa_{23} - c_{10} (\kappa_{13}^2 + \kappa_{23}^2) \right) 
\sin^2{\left( \frac{\Omega t}{2} \right)} \Big] \label{eq24c},
\end{align}
\end{subequations}
where, $\Omega=\sqrt{\Delta^2+\kappa_{13}^2+\kappa_{23}^2}$.
%================================================

\subsection{\texorpdfstring{$SU(3)$ Bloch vectors for $\Lambda$-configuration}{SU(3) Bloch vectors for Lambda configuration}}\label{subsec32}

\par 
Taking analogy from Eq.\eqref{eq1} of the qubit system \cite{Jaeger2006}, we define the density matrix operator for the three-level system in the SU(3) basis,
\begin{subequations}\label{eq25}
\begin{align}    
    \rho_{\text{TLS}}(t)
    &= \frac{1}{3}\left[\boldsymbol{\lambda}_0 + \frac{3}{2} \mathbf{n}_{T}(t) \cdot \boldsymbol{\lambda} \right] \nonumber \\ 
    &= \begin{bmatrix} 
        \frac{1}{6}(2 + 3 n_{3}(t) + \sqrt{3} n_{8}(t)) & \frac{1}{2}(n_{1}(t) - i n_{2}(t)) & \frac{1}{2}(n_{4}(t) - i n_{5}(t)) \\ 
        \frac{1}{2}(n_{1}(t) + i n_{2}(t)) & \frac{1}{6}(2 - 3 n_{3}(t) + \sqrt{3} n_{8}(t)) & \frac{1}{2}(n_{6}(t) - i n_{7}(t)) \\  
        \frac{1}{2}(n_{4}(t) + i n_{5}(t)) & \frac{1}{2}(n_{6}(t) + i n_{7}(t)) & \frac{1}{3}(1 - \sqrt{3} n_{8}(t))
    \end{bmatrix}, 
    \label{eq25a}
\end{align}
where $\mathbf{n}_{T}(t)$ is the SU(3) Bloch vector with the inner product defined as,  
\begin{align}
\mathbf{n}_{T}(t) \cdot \boldsymbol{\lambda} & ={n_{1}(t)} \lambda_1 +{n_{2}(t)} \lambda_2 + {n_{3}(t)}\lambda_3 + {n_{4}(t)}\lambda_4 \nonumber \\
&+ {n_{5}(t)}\lambda_5 + {n_{6}(t)}\lambda_6 + {n_{7}(t)}\lambda_7 + n_{8}(t)\lambda_8.  
\label{eq25b}
\end{align}
\end{subequations}
Here we emphasize that the judicious choice of $\frac{3}{2}$ in front of $\mathbf{n}_{T}(t) \cdot \boldsymbol{\lambda}$ in Eq.~\eqref{eq25a} rather than its other value discussed in the literature \cite{Goyal2016,Kurzy2016}, ensures that the density matrix $\rho_{\text{TLS}}$ gives precisely the same norm Eq.~\eqref{eq14} obtained for the qutrit system. This establishes the complete equivalence between the qutrit system with the three-level system. Furthermore, from Eq.~\eqref{eq25} it is easy to see that the {\it Purity} of such system can be related to the radius of the Bloch sphere \cite{Jaeger2006, wiki:Quantum_mechanics},
\begin{align}
Tr[\rho_{\text{TLS}}^2]=\frac{1}{3} \big(1+\frac{3}{2}|\bf{r}_{T}|^2\big), \quad \bf{r}_{T} \in \mathbb{R}^8. 
\label{eq26}
\end{align}
%Thus for the SU(3) Bloch sphere 
%ranges from $0 \leq |\mathbf{r}_{T}|^2 \leq \frac{4}{3}$ which 
and we have a bound, % on the \textit{Purity}, 
\begin{align}
\frac{1}{3} \leq Tr[\rho_{\text{TLS}}^2] \leq 1, \quad 0 \leq |\mathbf{r}_{T}|^2 \leq \frac{4}{3}
\label{eq27}
\end{align}
where the lower bound corresponds to the maximally mixed state.
\par 
To describe the dynamical behavior of the SU(3) Bloch vectors, we consider the wave function of the three-level system,\begin{align}
  \ket{\psi_{}^{\Lambda}(t)} = c_1(t)\ket{1} + c_2(t)\ket{2} + c_3(t)\ket{3}, 
  \label{eq28}
\end{align}
and corresponding density matrix in terms of the time dependent amplitudes are given by,
\begin{align}
    \rho_{\text{TLS}}(t)
     =\begin{bmatrix} c_1^{*}(t)c_1(t) & c_1^{*}(t)c_2(t) & c_1^{*}(t)c_3(t) \\ 
     c_2^{*}(t)c_1(t) & c_2^{*}(t)c_2(t) & c_2^{*}(t)c_3(t) \\ c_3^{*}(t)c_1(t) & c_3^{*}(t)c_2(t) & c_3^{*}(t)c_3(t) \end{bmatrix}.
     \label{eq29}
\end{align}
Comparing Eq.~\eqref{eq25} and Eq.\eqref{eq28}, we obtain the desired time-dependent SU(3) Bloch vectors of a three-level configuration,
\begin{align}
\mathbf{n}_T(t) =Tr\big[\boldsymbol{\lambda}\rho_{\text{TLS}}(t)\big], 
\label{eq30}
\end{align}
where its components are given by,
%\begin{subequations}
\begin{align} \label{eq31}
n_1(t) &=c_1^{*}(t) c_2(t)+c_1(t)c_2^{*}(t), \nonumber \\ 
n_2(t) &= i\big[c_1^{*}(t) c_2(t)-c_1(t)c_2^{*}(t)\big], \nonumber \\ 
n_3(t) &= c_1^{*}(t) c_1(t)-c_3(t)c_3^{*}(t), \nonumber  \\ 
n_4(t) &=  c_1^{*}(t) c_3(t)+c_1(t)c_3^{*}(t), \nonumber  \\ 
n_5(t) &= i[c_1^{*}(t) c_3(t)-c_1(t)c_3^{*}(t)], \\
n_6(t) &= c_2^{*}(t) c_3(t)+c_2(t)c_3^{*}(t), \nonumber \\
n_7(t) &= i[c_2^{*}(t) c_3(t)-c_2(t)c_3^{*}(t)], \nonumber \\
n_8(t) &= \frac{1}{\sqrt{2}}\big[c_1^{*}(t) c_1(t)+c_2^{*}(t)c_2(t)-2c_3^{*}(t)c_3(t)\big]. \nonumber 
\end{align}
For the $\Lambda$ configuration, substituting the amplitudes Eq.\eqref{eq24} into \eqref{eq31} we obtain the corresponding SU(3) Bloch vectors (For detail, see Appendix-A). Furthermore, it readily gives the norm of the seven-sphere $\mathbb{S}^7$ to be \cite{Sen2012},
\begin{subequations} \label{eq32}
\begin{align}
n_1^2(t) + n_2^2(t) + n_3^2(t) + n_4^2(t) + n_5^2(t) + n_6^2(t) + n_7^2(t) + n_8^2(t) = \frac{4}{3}.
\label{eq32a}
\end{align}
which exactly aligns with Eq.\eqref{eq14}, as derived in Section II through geometrical considerations. In addition, in this dynamical approach we note that at the zero detuning condition ($\Delta=0$), the seven-sphere $\mathbb{S}^7$ splits into two subspace: a four-sphere $\mathbb{S}^4$ with the norms  \cite{Sen2012}, 
\begin{align}
n_2^2(t) + n_3^2(t) &+ n_5^2(t) + n_6^2(t) + n_8^2(t) = \nonumber \\
&\frac{1}{3} \big[ -3 c_{10}^{*2} \left( c_{20}^2 + c_{30}^2 \right) + c_{20}^{*2} \left( 4 c_{20}^2 + 3 c_{30}^2 \right) \nonumber \\
&+ 2 c_{20} c_{20}^* c_{30} c_{30}^* + \left( 3 c_{20}^2 + 4 c_{30}^2 \right) c_{30}^{*2} + 2 c_{10} c_{10}^* \left( c_{20} c_{20}^* + c_{30} c_{30}^* \right) \nonumber \\
&+ c_{10}^2 \left( 4 c_{10}^{*2} - 3 \left( c_{20}^{*2} + c_{30}^{*2} \right) \right) \big], 
\label{eq32b}
\end{align}
and two-sphere $\mathbb{S}^2$,
\begin{align}
n_1^2(t) + n_4^2(t) + n_7^2(t) =& c_{10}^{*2} \left( c_{20}^2 + c_{30}^2 \right) - \left( c_{20}^* c_{30} - c_{20} c_{30}^* \right)^2 \nonumber \\ 
&+ 2 c_{10} c_{10}^*\left( c_{20} c_{20}^* + c_{30} c_{30}^* \right) + c_{10}^2 \left( c_{20}^{*2} + c_{30}^{*2} \right).
\label{eq32c}
\end{align}
\end{subequations}
respectively. The procedure outlined above can be easily extended for the $V$ and the $\Xi$ configurations \cite{Sen2012}. For the $V$ configuration, we have the following norms, 
\begin{subequations} \label{eq33}
\begin{align}
{n_1^V}^2(t)+{n_2^V}^2(t)+{n_3^V}^2(t)+{n_4^V}^2(t)
+{n_5^V}^2(t)+{n_6^V}^2(t)+{n_7^V}^2(t)+{n_8^V}^2(t) = \frac{4}{3}.
\label{eq33a}
\end{align}
Once again, at zero detuning condition ($\Delta^V=0$), the Bloch sphere split into two parts, 
%\begin{subequations}
\begin{align}
{n_1^V}^2+{n_3^V}^2&+{n_5^V}^2+{n_7^V}^2+{n_8^V}^2(t)=  \nonumber \\ 
& \frac{1}{3} \big[ 4 c_{20}^2 {c_{20}^{*}}^2 - 3 {c_{20}^{*}}^2 c_{30}^2 + 3 {c_{10}^{*}}^2 ( c_{20}^2 - c_{30}^2 ) \nonumber \\ 
&+ 2 c_{20} c_{20}^{*} c_{30} c_{30}^{*} - 3 c_{20}^2 {c_{30}^{*}}^2 + 4 c_{30}^2 {c_{30}^{*}}^2 \nonumber \\
&+ 2 c_{10} c_{10}^{*} ( c_{20} c_{20}^{*} + c_{30} c_{30}^{*}) + c_{10}^2 ( 4 {c_{10}^{*}}^2 + 3 {c_{20}^{*}}^2 - 3 {c_{30}^{*}}^2 )\big], 
\label{eq33b}
\end{align}
and, 
\begin{align}
{n_2^V}^2(t)+{n_4^V}^2(t)+{n_6^V}^2(t) & =  
{c_{10}^{*}}^2 \left( - c_{20}^2 + c_{30}^2 \right) + \left( c_{20}^{*} c_{30} + c_{20} c_{30}^{*} \right)^2 + \nonumber \\ 
&2 c_{10} c_{10}^{*} \left( c_{20} c_{20}^{*} + c_{30} c_{30}^{*} \right) + c_{10}^2 ( - {c_{20}^{*}}^2 + {c_{30}^{*}}^2).
\label{eq33c}
\end{align}
\end{subequations}
Similarly for the $\Xi$ configuration we have,
\begin{subequations} \label{eq34}
\begin{align}
{n_1^{\Xi}}^2(t)+{n_2^{\Xi}}^2(t)+{n_3^{\Xi}}^2(t)+{n_4^{\Xi}}^2(t)
+{n_5^{\Xi}}^2(t)+{n_6^{\Xi}}^2(t)+{n_7^{\Xi}}^2(t)+{n_8^{\Xi}}^2(t) = \frac{4}{3},
\label{eq34a}
\end{align}
and at resonance ($\Delta^{\Xi}=0$) we have, 
%\begin{subequations}
\begin{align}
{n_2^{\Xi}}^2(t)+{n_3^{\Xi}}^2(t)&+{n_4^{\Xi}}^2(t)+{n_{7}^{\Xi}}^2(t)+{n_8^{\Xi}}^2(t) = \nonumber \\ 
&\frac{1}{3} \big[4 c_{20}^2 {c_{20}^{*}}^2 - 3 {c_{20}^{*}}^2 c_{30}^2- 3 {c_{10}^{*}}^2 (c_{20}^2 - c_{30}^2), \nonumber\\
& + 2 c_{20} c_{20}^{*} c_{30} c_{30}^{*} - 3 c_{20}^2 {c_{30}^{*}}^2 + 4 c_{30}^2 {c_{30}^{*}}^2 + 2 c_{10} c_{10}^{*} \left(c_{20} c_{20}^{*} + c_{30} c_{30}^{*}\right) \nonumber \\
&+ c_{10}^2 (4 {c_{10}^{*}}^2 - 3 {c_{20}^{*}}^2 + 3 {c_{30}^{*}}^2) \big], 
\label{eq34b}
\end{align}
and 
\begin{align}
{n_1^{\Xi}}^2(t)+{n_5^{\Xi}}^2(t)+{n_6^{\Xi}}^2(t) &= 
{c_{10}^{*}}^2 \left(c_{20}^2 - c_{30}^2\right) + (c_{20}^{*} c_{30} + c_{20} c_{30}^{*})^2 \nonumber \\
& + 2 c_{10} c^{*}_{10} \left(c_{20} c_{20}^{*} + c_{30} c_{30}^{*}\right) + c_{10}^2 ({c_{20}^{*}}^2 - {c_{30}^{*}}^2),
\label{eq34c}
\end{align}
\end{subequations}
respectively. 
\section{Trajectories of the SU(3) Bloch vectors} \label{section4}
\par 
We are now in a position to discuss the phase portraits of  different configurations of the three-level system to examine the dynamics of the SU(3) Bloch vectors. For numerical studies, we work with the initial condition where all three levels are equally populated, i.e., $c_{10} = c_{20} = c_{30} = 1/\sqrt{3}$ at $t=0$. For example, for $\Lambda$ configuration, the norm of the four-sphere $\mathbb{S}^4$ at zero detuning is given by, 
\begin{subequations}\label{eq35}
\begin{align}
n_2^2(t)+n_3^2(t)+n_5^2(t)+n_6^2(t)+n_8^2(t) &=\frac{4}{9},
\label{eq35a}
\end{align}
from Eq.(32b) and that for the two-sphere $\mathbb{S}^2$ we have, 
\begin{align}
n_1^2(t)+n_4^2(t)+n_7^2(t) &= \frac{8}{9},  %\quad (> 1)
\label{eq35b}
\end{align}
\end{subequations}
from Eq.(32c).
\par 
Let us begin by considering the time-series plots of all Bloch vectors at resonance ($\Delta=0$) and off-resonance ($\Delta \neq 0$), as shown in Figs.1-3 for the $\Lambda$, $V$ and $\Xi$ configurations. The periodic behavior of the evolution curves at resonance ($\Delta=0$) is clearly evident from Figs.(1a,2a,3a). We note that this periodicity changes its pattern for the nonzero value ($\Delta \neq 0$) of the detuning parameter shown in Figs.(1b,2b,3b).
\par 
To make the physical content of the dynamics more transparent, we now consider the phase space plots (i.e., $n_i(t)$ versus $\dot{n}_i(t)$ plots) of the Bloch vectors shown in Fig.4-6 for three configurations (For a detailed derivation of $n_i(t)$ and $\dot{n}_i(t)$, see Appendix A and B). At resonance, the phase portraits of three configurations reveal closed curves, signifying stable and periodic behavior of the Bloch trajectories. However, when a finite detuning is turned on, several Bloch vectors exhibit a noticeable precession, contrast to those seen at resonance. Thus it is worth noting that at zero detuning, the entire SU(3) Bloch sphere $\mathbb{S}^7$ splits as into two sectors, $\mathbb{S}^4$ and $\mathbb{S}^2$ which are characterized by their distinctive Bloch trajectories.
%\par 
In Figures 4a,5a, and 6a, we have displayed the phase portrait of the $\mathbb{S}^4$ sector for the $\Lambda$ configuration ($i = 2,3,5,6,8$), for the $V$ configuration ($i=1,3,5,7,8$) and for the $\Xi$ configuration ($i=2,3,4,7,8$), respectively. In particular we note that all configurations are closed and exhibit complex trisectrix family of curves with their characteristic looping and branching patterns. Such structure is highly sensitive to the system's parameters revealing the richness of underlying geometry and nontrivial topology of that sector. In Figures 4b,5b, and 6b, display the phase portraits of the $\mathbb{S}^2$ sector for the $\Lambda$ configuration ($i = 1,4,7$), the $V$ configuration ($i = 2,4,6$) and the $\Xi$  configurations ($i = 1,5,6$). In contrast to the previous sector, these plots for all configurations exhibits a trajectory that is closer to the ellipse at resonance and display a regular precessional feature at off-resonance.  
%================================

\section{Conclusions}\label{section7}
\par 
In this paper, we have investigated the Bloch space of the qutrit system by studying the Bloch trajectories for different configurations of three-level system. From the geometric perspective, when such a system is parameterized with spherical polar coordinates, it leads to a Bloch seven-sphere  
$\mathbb{S}^7$ with the norm $4/3$ which exceeds 1. This parameterization further facilitates the existence of three distinct principal cardinal states along with their superposed states. We then proceed to evaluate the norm for each configuration from the dynamical consideration and show that it is in conformity with the norm obtained by the aforesaid geometric approach. 
%\par 
In this approach, additionally, the Bloch seven-sphere $\mathbb{S}^7$ is found to split into a four-sphere $\mathbb{S}^4$ and a two-sphere $\mathbb{S}^2$ at resonance. We have compared the time series and phase space plots of the SU(3) Bloch vectors of all three-level configurations at and off-resonance. For $\mathbb{S}^4$ sector, the Bloch trajectory reveals closed but a complex, trisectrix class of closed orbits, while for the $\mathbb{S}^2$ sector, we have approximately elliptical trajectory. On the other hand, at off-resonance, such splitting vanishes and all the orbits display a precessional motion. Such distinctive dynamics of the Bloch vectors is reflecting an intricate and nontrivial geometry of the Bloch space of the qutrit system which requires further exploration. The analysis presented here is limited to a narrow range of numerical parameters, but extending it to a broader parameter range could uncover additional dynamical features of high-dimensional quantum systems, which might have deeper geometrical implications.

\vfill
\section*{Acknowledgment}
 
We thank Dr Sovik Roy for discussion on the properties of the trisectrix curves. 
%======================================

%%%%%%%%%%%%%%%%%%%%%%%%%%%%%%%%%%%%%
\newpage
%% Figures & Subfigures

\section*{Figure Captions: }

%\section*{Figure Caption: 1 Lambda}

\begin{figure}[ht] % Start of the figure environment
    \centering
    % Subfigure 1a
    \begin{subfigure}{.9\textwidth} % Width of each subfigure
        \centering
        \includegraphics[width=.9\textwidth]{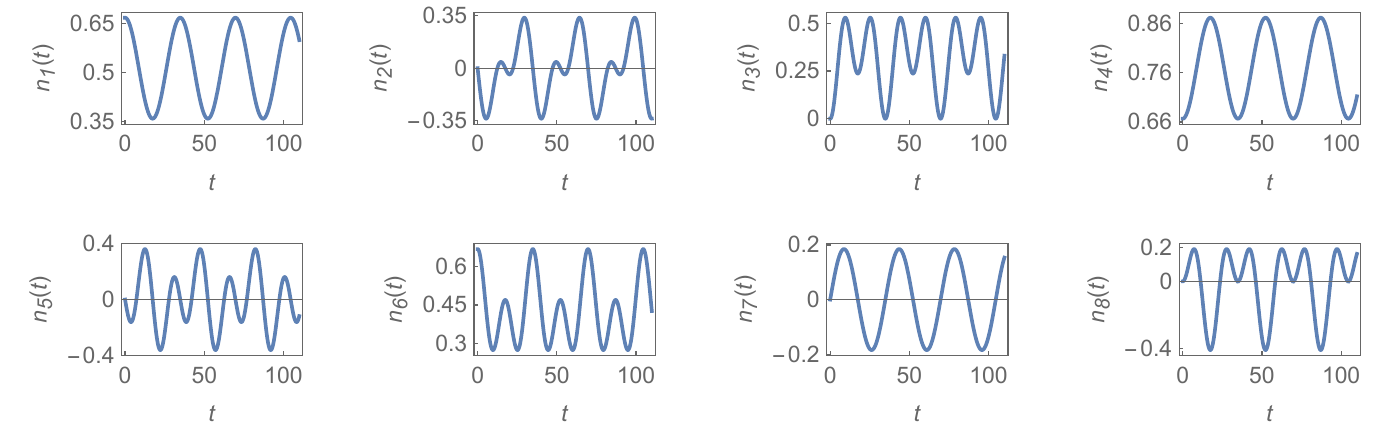} % Replace 'image1' with your image filename
        \caption*{Fig1a: Time evolution of the SU(3) Bloch vectors $n_i(t)$ at resonance ($\Delta=0$) for the $\Lambda$ configuration with $\kappa_{13}=.3$, $\kappa_{23}=.2$ and $c_1(0)=c_2(0)=c_3(0)=1/\sqrt{3}$.} % Subfigure caption
        \label{Lambdafig:1a}  % Label for referencing
    \end{subfigure}
%    \hfill % Space between the subfigures
    
    % Subfigure 2b
    \begin{subfigure}{.9\textwidth}
        \centering
        \includegraphics[width=.9\textwidth]{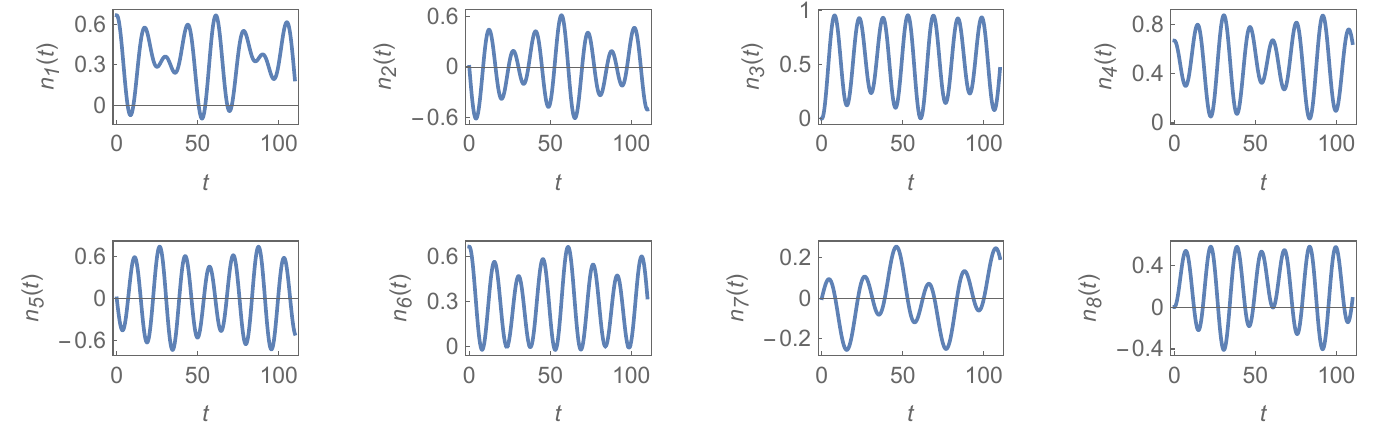} % Replace 'image2' with your image filename
        \caption*{Fig1b: Time evolution of the SU(3) Bloch vectors at off-resonance ($\Delta=.2$) for the $\Lambda$ configuration with the same parameter values. The pattern of the periodic evolution of the Bloch vectors is found to be altered.}
        \label{Lambdafig:1b}
    \end{subfigure}
%    \hfill % Space between the subfigures
%        \caption*{Fig1: Time evolution of the Bloch vectors for the $\Lambda$ configuration.} % Overall figure caption
%    \label{fig:main}  % Label for the main figure
\end{figure}

%%%%%%%%%%%%%%%%%%%%%%%%%%%%%%
\newpage
%\section*{Figure Caption: 2 Vee}

\begin{figure}[ht] % Start of the figure environment 
    \centering
    % Subfigure 2a
    \begin{subfigure}{.9\textwidth} % Width of each subfigure
        \centering
        \includegraphics[width=.9\textwidth]{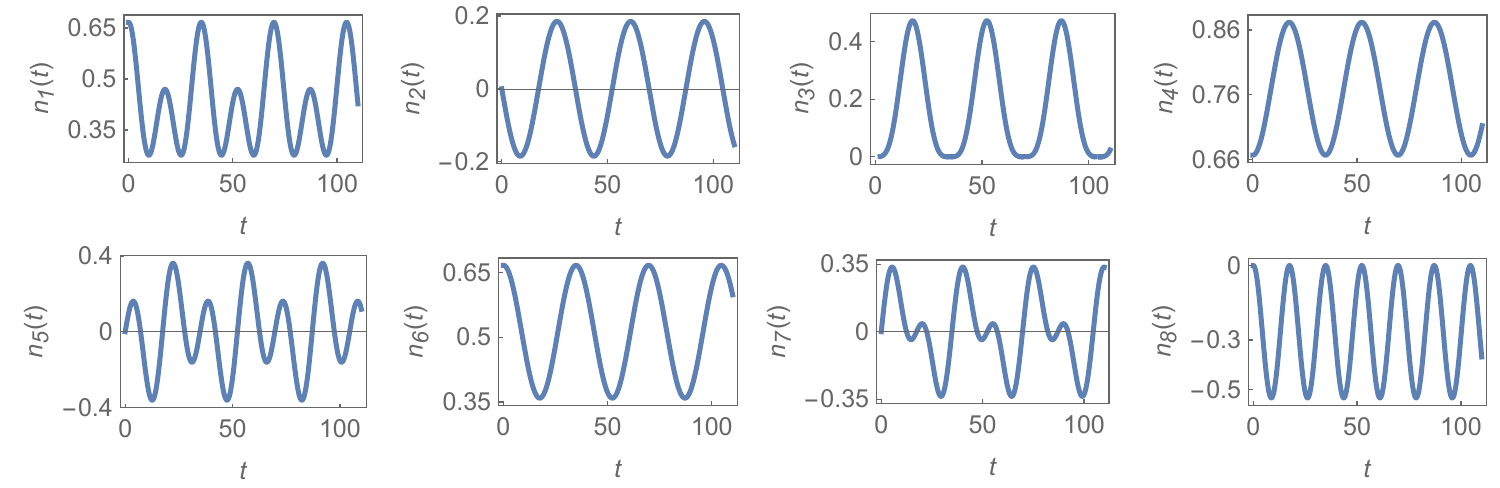} % Replace 'image1' with your image filename
        \caption*{Fig2a: Time evolution of the SU(3) Bloch vectors $n_i(t)$ at resonance ($\Delta=0$) for the $V$ configuration with $\kappa_{13}=.3$, $\kappa_{12}=.2$ and $c_1(0)=c_2(0)=c_3(0)=1/\sqrt{3}$.} % Subfigure caption
        \label{Veefig:2a}  % Label for referencing
    \end{subfigure}
%    \hfill % Space between the subfigures
    
    % Subfigure 2b
    \begin{subfigure}{.9\textwidth}
        \centering
        \includegraphics[width=.9\textwidth]{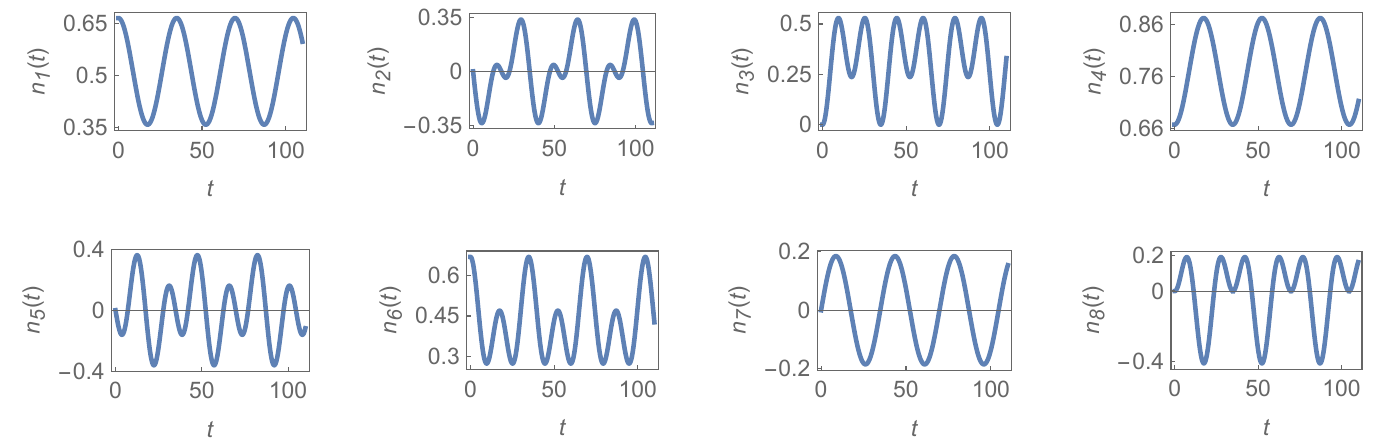} % Replace 'image2' with your image filename
        \caption*{Fig2b: The time evolution of the Bloch vectors for the $V$ configuration at off-resonance ($\Delta = 0.2$) with same values of the parameters displays distinct pattern of  periodicity.}
        \label{Veefig:2b}
    \end{subfigure}
%    \hfill % Space between the subfigures
%        \caption*{Fig2: Time evolution of the Bloch vectors for the $V$ configuration.} % Overall figure caption
%    \label{Vee:2b}  % Label for the main figure
\end{figure}

%%%%%%%%%%%%%%%%%%%%%
\newpage 
%\section*{Figure Caption: 3 Cascade }

\begin{figure}[ht] % Start of the figure environment 
    \centering
    % Subfigure 1
    \begin{subfigure}{.9\textwidth} % Width of each subfigure
        \centering
        \includegraphics[width=.9\textwidth]{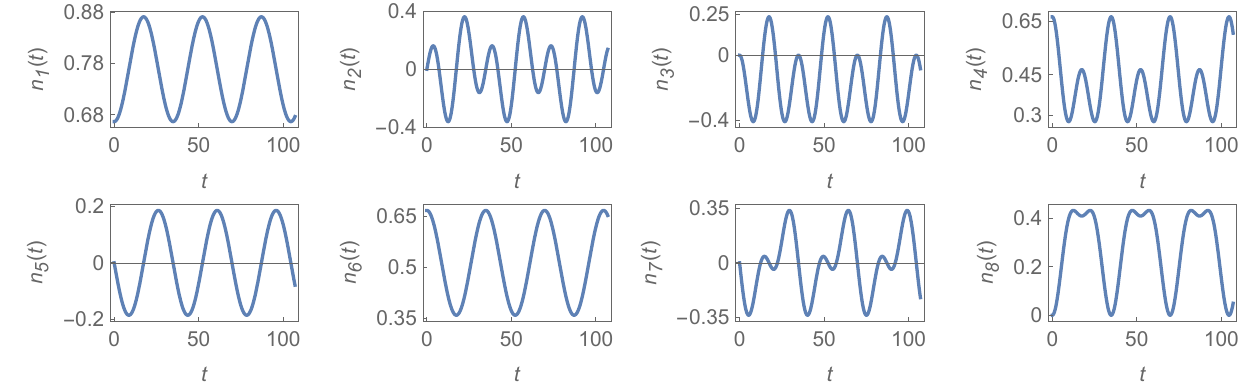} % Replace 'image1' with your image filename
        \caption*{Fig3a: Time evolution of the Bloch vectors $n_i(t)$ at resonance ($\Delta=0$) for the $\Xi$ configuration with $\kappa_{12}=.2$, $\kappa_{23}=.3$,  and $c_1(0)=c_2(0)=c_3(0)=1/\sqrt{3}$.} % Subfigure caption
        \label{Xifig:3a}  % Label for referencing
    \end{subfigure}
%    \hfill % Space between the subfigures
\end{figure}
    
\begin{figure}[ht] % Start of the figure environment 
    \centering
    % Subfigure 1
    \begin{subfigure}{.9\textwidth} % Width of each subfigure
        \centering
        \includegraphics[width=.9\textwidth]{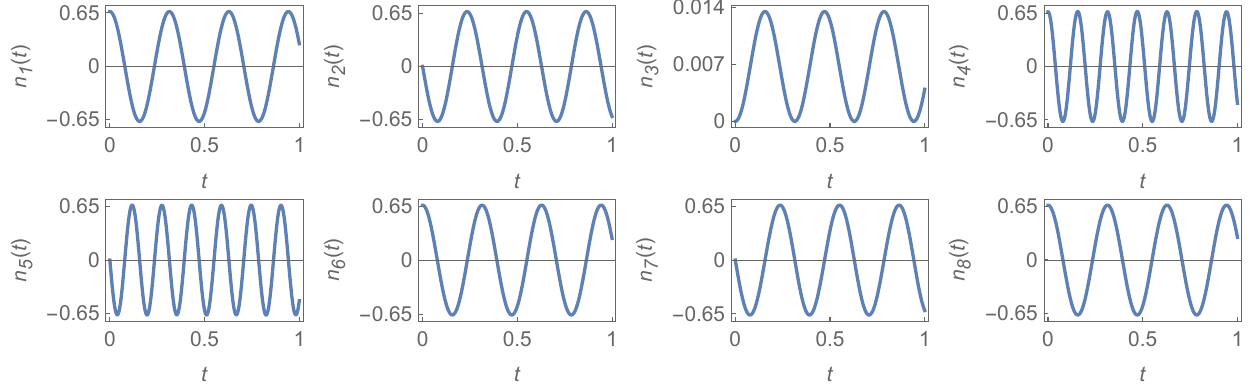} % Replace 'image1' with your image filename
        \caption*{Fig3b: Time evolution of the SU(3) Bloch vectors at off-resonance for the $\Xi$ configuration with same values of above parameters with detuning $\Delta=20$ showing distinct pattern of periodicity.} % Subfigure caption
        \label{Xifig:3b}  % Label for referencing
    \end{subfigure}
%    \hfill % Space between the subfigures   
%        \caption*{Fig3: Time evolution of the Bloch vectors for the $\Xi$ configuration.} % Overall figure caption
%    \label{Xifig:3b}  % Label for the main figure
\end{figure}

%%%%%%%%%%%%%%%%%%%%%%
\newpage 
%\section*{Figure Caption: 4a Lambda} 

\begin{figure}[ht] % Start of the figure environment 
    \centering
    % Subfigure 2
    \begin{subfigure}{.9\textwidth} % Width of each subfigure
        \centering
        \includegraphics[width=.9\textwidth]{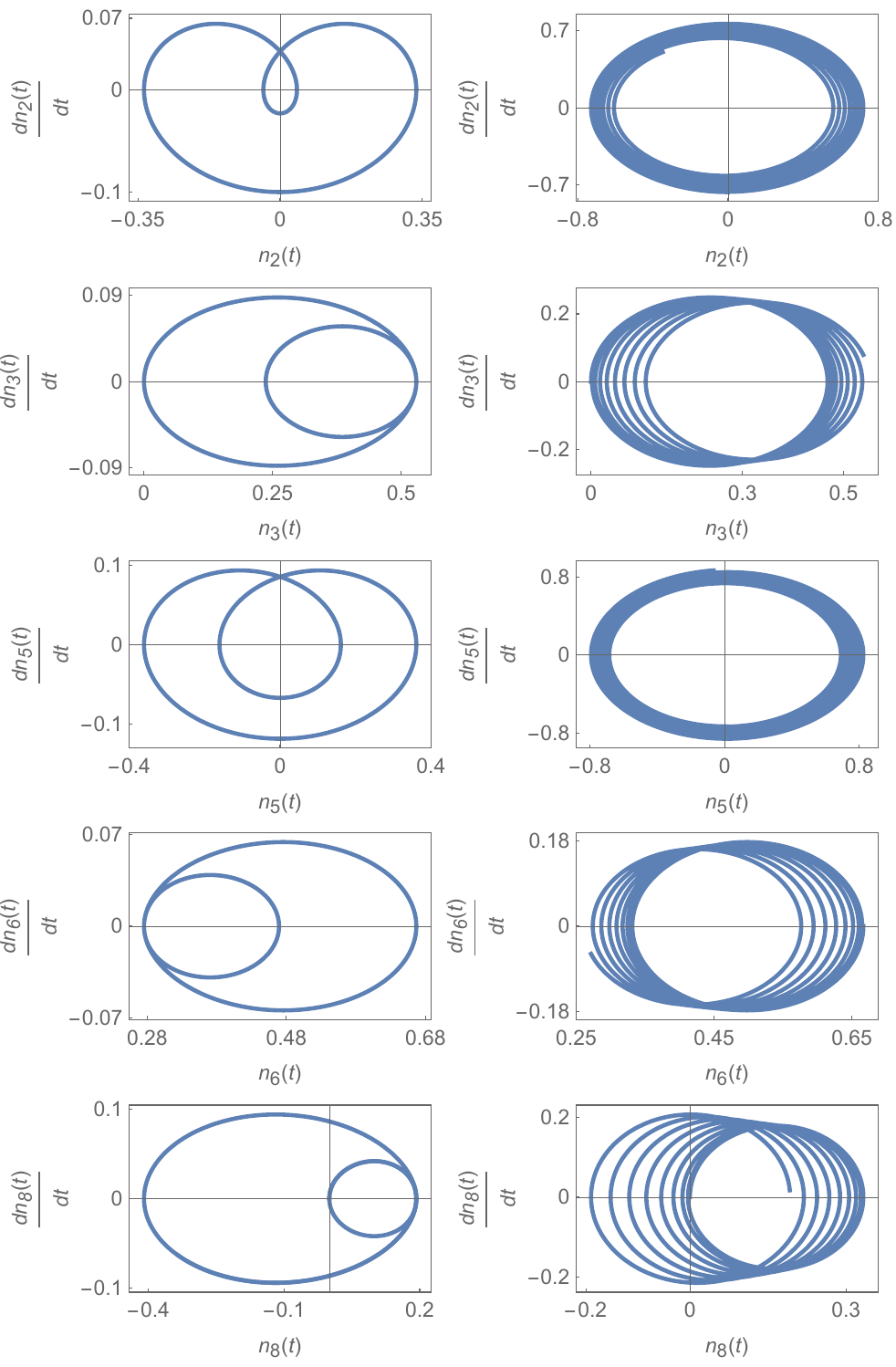} 
        \caption*{Fig.4a: Phase portraits of the SU(3) Bloch vectors $n_i(t)$ ($i=2,3,5,6,8$) for the $4$-sphere $\mathbb{S}^4$ sector of the $\Lambda$ configuration at resonance ($\Delta=0$, Left column) are compared with those at off-resonance ($\Delta=1.2$, Right column) with the parameters, $\kappa_{13}=.3$, $\kappa_{23}=.2$ and $c_1(0)=c_2(0)=c_3(0)=1/\sqrt{3}$. The precession of the Bloch trajectories is clearly evident.} % Subfigure caption
        \label{Lambda5fig:4a}  % Label for referencing
    \end{subfigure}
    \hfill % Space between the subfigures
\end{figure}

%%%%%%%%%%%%%%%%%%%%%%%%%%%%%%
\newpage 
%\section*{Figure Caption4b: Lambda}
    
\begin{figure}[ht] % Start of the figure environment 
    \centering
    % Subfigure 2
    \begin{subfigure}{0.9\textwidth} % Width of each subfigure
        \centering
        \includegraphics[width=.9\textwidth]{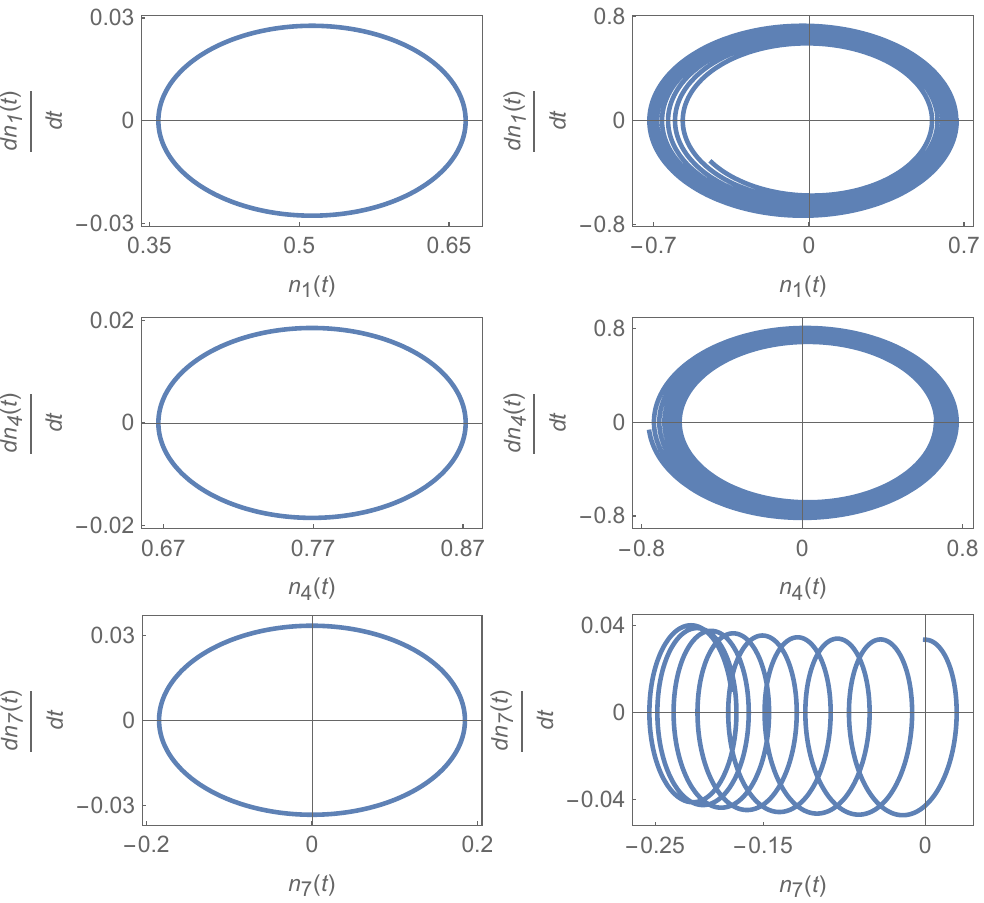} % Replace 'image1' with your image filename
        \caption*{Fig.4b: The phase portrait of the SU(3) Bloch vectors $n_i(t)$ ($i=1,4,7$) for the $2$-sphere $\mathbb{S}^2$ sector of the $\Lambda$ configuration at resonance ($\Delta=0$, Left column) and off-resonance ($\Delta=1.2$, Right column) are compared using aforesaid  parameters values.} % Subfigure caption
        \label{Lambda3fig:4b}  % Label for referencing
    \end{subfigure}
    \hfill % Space between the subfigures
\end{figure}

%%%%%%%%%%%%%%%%%%%%%%%%%%%%
\newpage 
%\section*{Figure Caption: 5a Vee }

\begin{figure}[ht] % Start of the figure environment 
    \centering
    % Subfigure 1
    \begin{subfigure}{0.9\textwidth} % Width of each subfigure
        \centering
        \includegraphics[width=.9\textwidth]{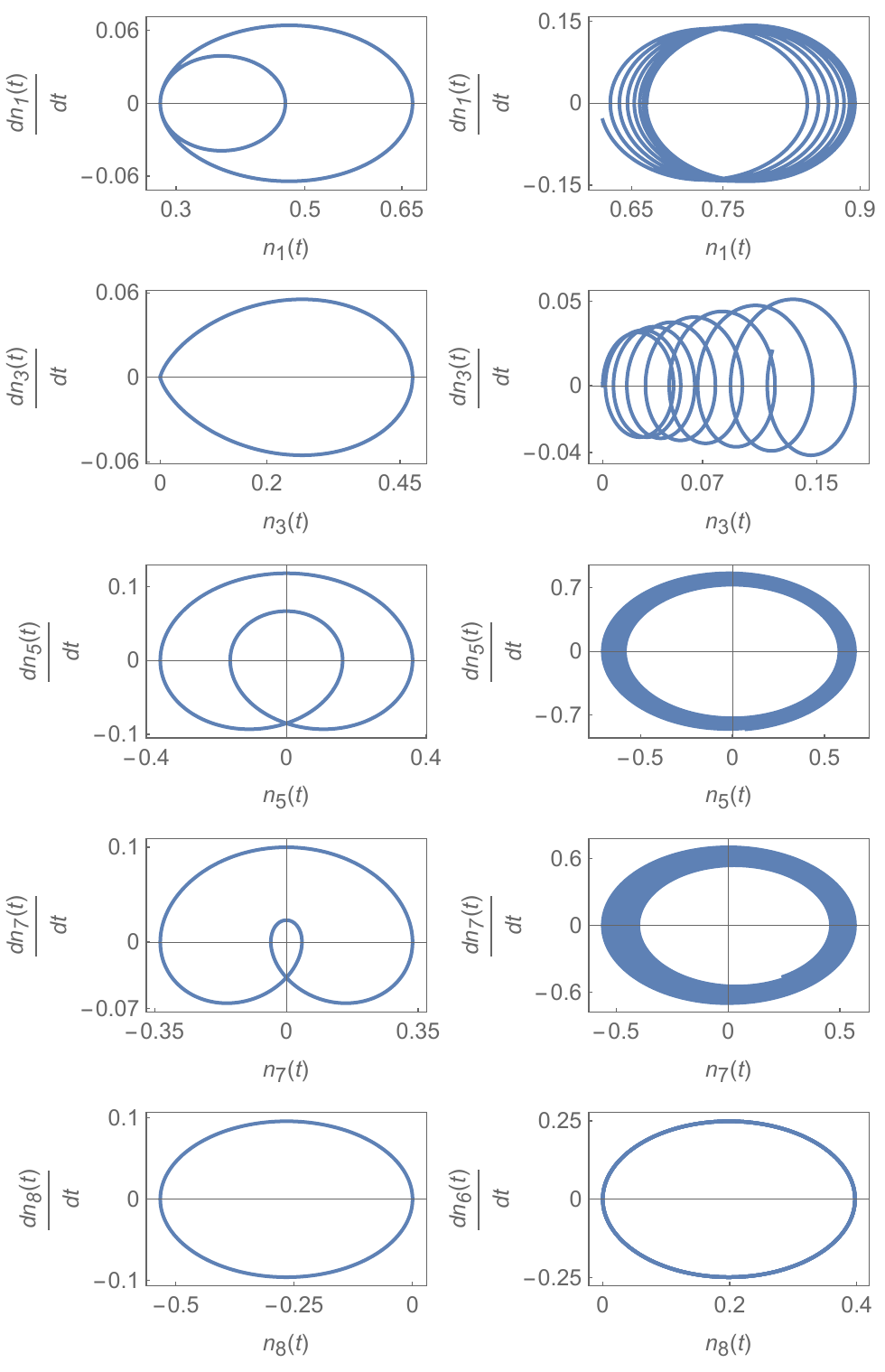} % Replace 'image1' with your image filename
        \caption*{Fig.5a: Phase portraits of the SU(3) Bloch vectors $n_i(t)$ ($i=1,3,5,7,8$) for the $4$-sphere $\mathbb{S}^4$ sector of the $V$ configuration at resonance ($\Delta=0$, Left column) are compared with those at off-resonance($\Delta=0.2$, Right column) with the parameters, $\kappa_{13}=.3$, $\kappa_{12}=.2$ and $c_1(0)=c_2(0)=c_3(0)=1/\sqrt{3}$. The precession of the Bloch trajectories is clearly evident.} % Subfigure caption
        \label{Vee5fig:5a}  % Label for referencing
    \end{subfigure}
    \hfill % Space between the subfigures
\end{figure}

%%%%%%%%%%%%%%%%%%%%%%%%%
\newpage 
%\section*{Figure Caption5b: Vee }
    
\begin{figure}[ht] % Start of the figure environment VEE
    \centering
    % Subfigure 9
    \begin{subfigure}{0.8\textwidth} % Width of each subfigure
        \centering
        \includegraphics[width=.9\textwidth]{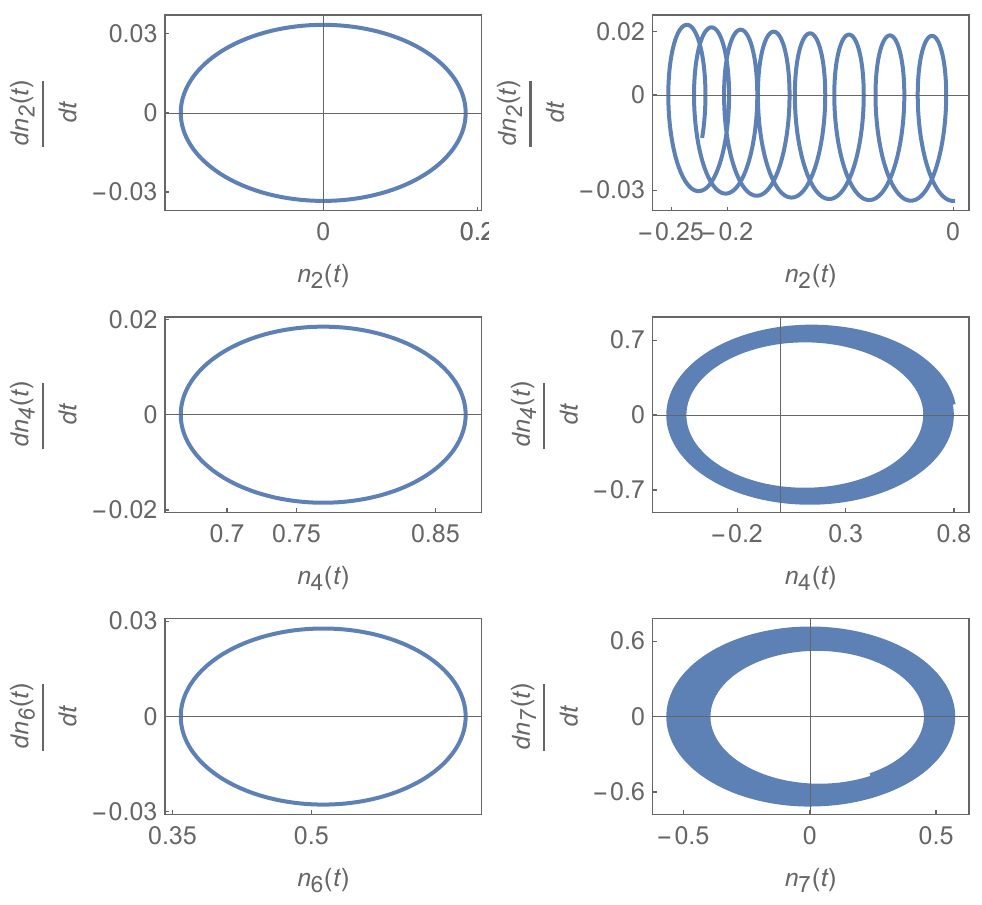} % Replace 'image1' with your image filename
        \caption*{Fig.5b: The phase portrait of the SU(3) Bloch vectors $n_i(t)$ ($i=2,4,6$)  for the $2$-sphere $\mathbb{S}^2$ sector of the $V$ configuration at resonance ($\Delta=0$, Left column) and off-resonance ($\Delta=0.2$, Right column) using the same parameter values.} % Subfigure caption
        \label{Vee3fig:5b}  % Label for referencing
    \end{subfigure}
    \hfill % Space between the subfigures
\end{figure}

%%%%%%%%%%%%%%%%%%%%%%%%%%%%%%%%%%%%%%%%%%%%%%%
\newpage 
%\section*{Figure Caption: 6a Cascade }

\begin{figure}[ht] % Start of the figure environment 
    \centering
    % Subfigure 1
    \begin{subfigure}{0.9\textwidth} % Width of each subfigure
        \centering
        \includegraphics[width=.9\textwidth]{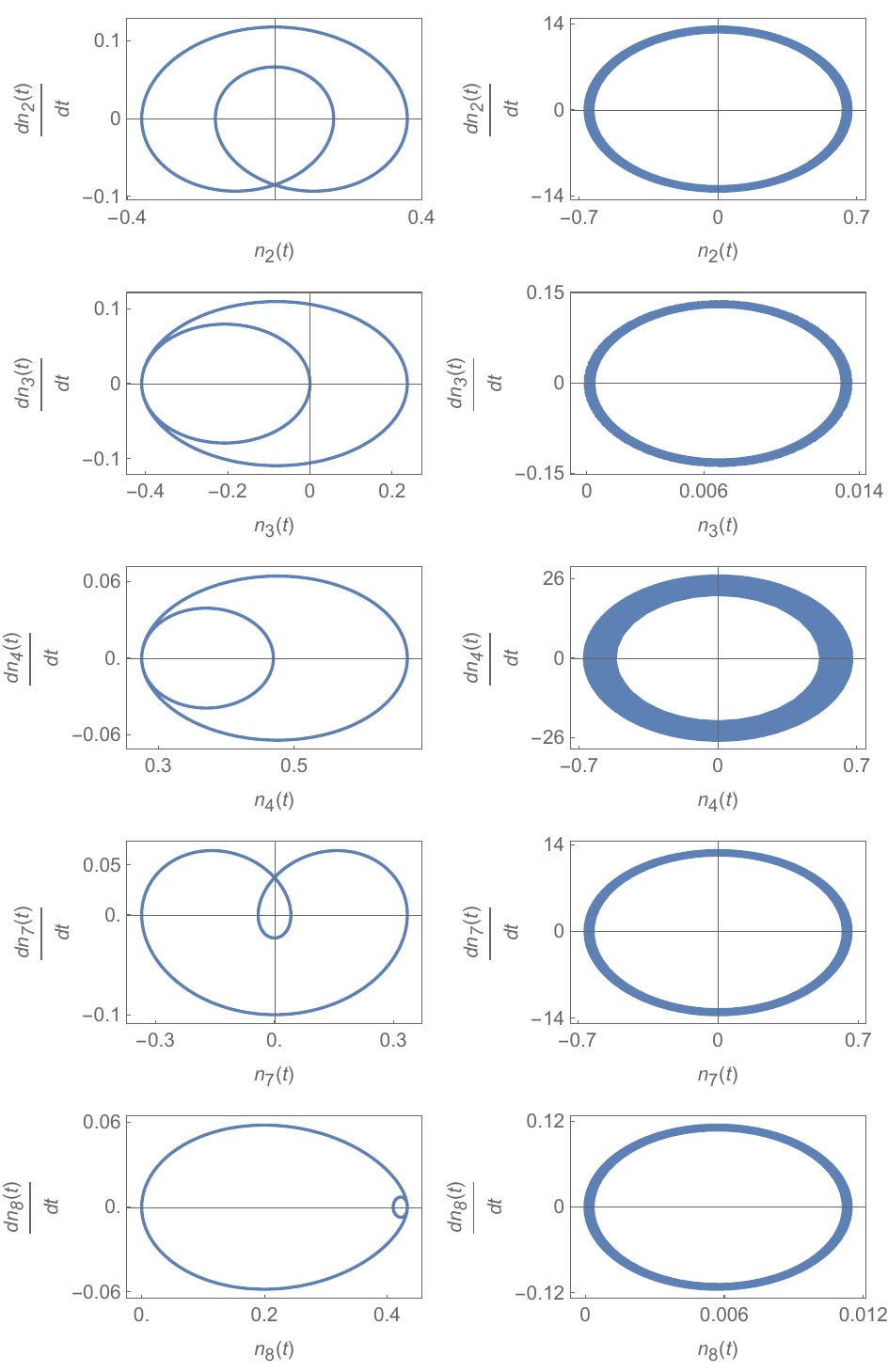} % Replace 'image1' with your image filename
        \caption*{Fig.6a: Phase portrait of the SU(3) Bloch vectors $n_i(t)$ ($i=2,3,5,6,8$) for the $4$-sphere $\mathbb{S}^4$ sector of the $\Xi$ configuration at resonance ($\Delta=0$, Left column) is compared with those at off-resonance ($\Delta=20$, Right column) with $\kappa_{12}=.2$, $\kappa_{23}=.3$ and $c_1(0)=c_2(0)=c_3(0)=1/\sqrt{3}$. Like previous two configurations, the Bloch trajectories exhibit a  precession for the non-zero value of detuning.} % Subfigure caption
        \label{Xi5fig:6a}  % Label for referencing
    \end{subfigure}
    \hfill % Space between the subfigures
\end{figure}

%%%%%%%%%%%%%%%%%%%%%%%%%%%%%
\newpage 
%\section*{Figure Caption: 6b Cascade }

\begin{figure}[ht] % Start of the figure environment 
    \centering
    % Subfigure 2
    \begin{subfigure}{0.9\textwidth} % Width of each subfigure
        \centering
        \includegraphics[width=.9\textwidth]{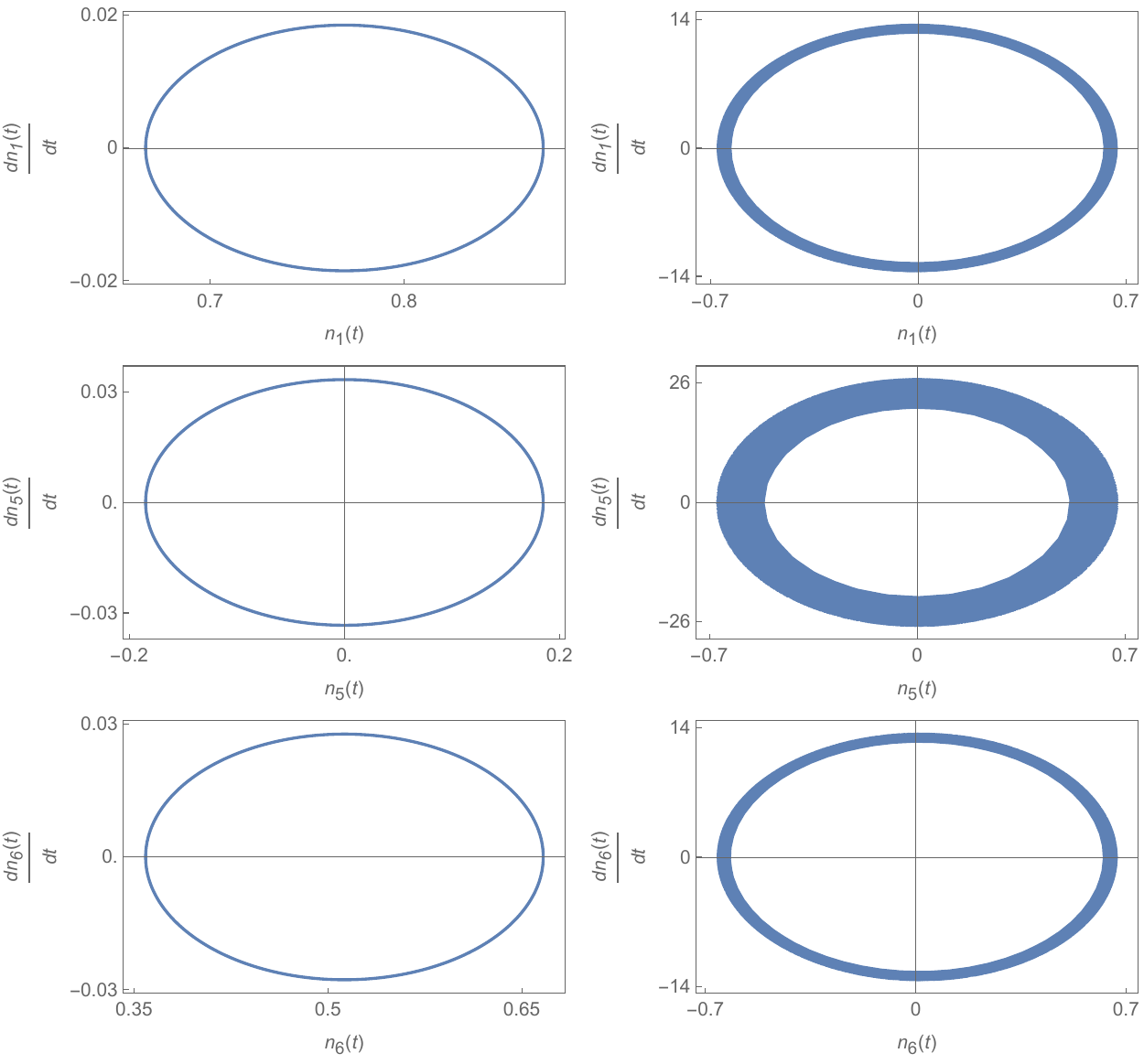} % Replace 'image1' with your image filename
        \caption*{Fig.6b: The phase portrait of the SU(3) Bloch vectors $n_i(t)$ ($i=1, 5, 6$) for the $2$-sphere $\mathbb{S}^2$ sector of the $\Xi$ configuration at resonance ($\Delta=0$, Left column) and off-resonance ($\Delta=20$, Right column) which shows a precession at detuning $\Delta=20$, using the same parameter values.} % Subfigure caption2
        \label{Xi3fig:6b}  
    \end{subfigure}
    \hfill % Space between the subfigures
\end{figure}

%===================================
%%%%%%%%%%%%%%%%%%%%%%%%%%%%%%%%%%%%%%%%%%%
\clearpage 
%%%%%%%%%%%%%%%%%%%%%%%%%%%%%%%%
\newpage 
\appendix
%\section{Bloch vectors for the $\Lambda$ configuration:}\label{secA}
\section*{Appendix A: SU(3) Bloch vectors for the \texorpdfstring{$\Lambda$}{Lambda} configuration}\label{secA}

% Redefine equation numbering for Appendix
\renewcommand{\theequation}{A.\arabic{equation}}
\setcounter{equation}{0} % Reset equation counter

\par 
The phase portrait of a given configurations requires the plotting of the phase plane coordinates 
\big($n_i(t),\dot{n}_i(t)$\big). In this Appendix we have explicitly displayed the SU(3) Bloch vectors $n_i(t)$ for the $\Lambda$ configuration by plugging in the amplitudes from Eq.\eqref{eq24} into Eq.\eqref{eq31}:

\begin{align}
n_1(t) &= \frac{e^{-i\frac{1}{2}\Delta t}}{(\kappa_{13}^2 + \kappa_{23}^2)\Omega^2} \Big\{
    e^{i\frac{1}{2}\Delta t}\kappa_{23} \Big[ c_{10}(c_{10}^* \Delta + c_{30}^* \kappa_{13} + c_{20}^* \kappa_{23}) (\kappa_{13}^2 + \kappa_{23}^2) \nonumber\\
    & \quad + (c_{30} \kappa_{13} + c_{20} \kappa_{23})(-c_{30}^* \Delta \kappa_{13} - c_{20}^* \Delta \kappa_{23} + c_{10}^*(\kappa_{13}^2 + \kappa_{23}^2))\Big] \nonumber\\
    & \quad + \kappa_{13}(c_{20}^* \kappa_{13} - c_{30}^* \kappa_{23}) \Omega \Big[c_{10} \Omega \cos\left(\frac{\Omega t}{2}\right) - i(c_{10} \Delta + c_{30} \kappa_{13} + c_{20} \kappa_{23}) \sin\left(\frac{\Omega t}{2}\right)\Big] \nonumber\\
    &+e^{i \Delta t} \kappa_{13}(c_{20} \kappa_{13} - c_{30} \kappa_{23}) \Omega \Big[c_{10}^* \Omega \cos\left(\frac{\Omega t}{2}\right) + i(c_{10}^* \Delta + c_{30}^* \kappa_{13} + c_{20}^* \kappa_{23}) \sin\left(\frac{\Omega t}{2}\right)\Big] \nonumber\\
    & \quad + e^{\frac{i \Delta t}{2}} \Delta \kappa_{23} \Big[\Big((c_{10} \Delta + c_{30} \kappa_{13} + c_{20} \kappa_{23})(c_{30}^* \kappa_{13} + c_{20}^* \kappa_{23}) \nonumber\\
    & \quad + c_{10}^*(c_{30} \Delta \kappa_{13} + c_{20} \Delta \kappa_{23} - c_{10}(\kappa_{13}^2 + \kappa_{23}^2))\Big) \cos\Omega t \nonumber\\
    & \quad + i(c_{10}^*(c_{30} \kappa_{13} + c_{20} \kappa_{23}) - c_{10}(c_{30}^* \kappa_{13} + c_{20}^* \kappa_{23})) \Omega \sin\Omega t\Big]\Big\}, 
    \label{A1}
\end{align}

\begin{align}
n_2^\Lambda(t) & = \frac{1}{(\kappa_{13}^2 + \kappa_{23}^2)\Omega} \Big\{
    i e^{-\frac{i \Delta t}{2}} \kappa_{13} \Big[ c_{10}^* e^{i \Delta t} (c_{20} \kappa_{13} - c_{30} \kappa_{23}) \nonumber\\
    & \quad + c_{10} (-c_{20}^* \kappa_{13} + c_{30}^* \kappa_{23}) \Big] \Omega \cos\left(\frac{\Omega t}{2}\right) \nonumber\\
    & \quad + \kappa_{23} \Big[c_{10}^* (c_{30} \kappa_{13} + c_{20} \kappa_{23}) - c_{10} (c_{30}^* \kappa_{13} + c_{20}^* \kappa_{23}) \Big] \Omega \cos\Omega t \nonumber\\
    & \quad + i e^{-\frac{i \Delta t}{2}} \kappa_{13} \Big[ e^{i \Delta t} (c_{10}^* \Delta + c_{30}^* \kappa_{13} + c_{20}^* \kappa_{23}) (c_{20} \kappa_{13} -   
   c_{30} \kappa_{23}) \nonumber\\
    & \quad + (c_{10} \Delta + c_{30} \kappa_{13} + c_{20} \kappa_{23}) (c_{20}^* \kappa_{13} - c_{30}^* \kappa_{23}) \Big] \sin\left(\frac{\Omega t}{2}\right)    
     \nonumber\\
    & \quad + i \kappa_{23} \Big[ (c_{10} \Delta + c_{30} \kappa_{13} + c_{20} \kappa_{23}) (c_{30}^* \kappa_{13} + c_{20}^* \kappa_{23}) \nonumber\\
    & \quad + c_{10}^* \left(c_{30} \Delta \kappa_{13} + c_{20} \Delta \kappa_{23} - c_{10} (\kappa_{13}^2 + \kappa_{23}^2)\right) \Big] \sin\Omega t
     \Big\}, 
\end{align}

\begin{align}
%###############################################################
n_3^\Lambda(t)&=  \frac{1}{2 (\kappa_{13}^2 + \kappa_{23}^2)^2 \Omega^2} \Big\{
    \kappa_{13}^4 \left(c_{30} \kappa_{13} (c_{10}^* \Delta + c_{30}^* \kappa_{13}) - 2 c_{20} c_{20}^* (\Delta^2 + \kappa_{13}^2)\right) \nonumber\\
    & \quad + \kappa_{13}^3 \left(2 (c_{20}^* c_{30} + c_{20} c_{30}^*) \Delta^2 + c_{10}^* c_{20} \Delta \kappa_{13} + 3 (c_{20}^* c_{30} + c_{20} c_{30}^*) \kappa_{13}^2\right) \kappa_{23} \nonumber\\
    & \quad - \kappa_{13}^2 \left(4 c_{30} c_{30}^* \Delta^2 - 3 c_{10}^* c_{30} \Delta \kappa_{13} + (c_{20} c_{20}^* + c_{30} c_{30}^*) \kappa_{13}^2\right) \kappa_{23}^2 \nonumber\\
    & \quad + \kappa_{13} \left(-2 (c_{20}^* c_{30} + c_{20} c_{30}^*) \Delta^2 + 3 c_{10}^* c_{20} \Delta \kappa_{13} + 3 (c_{20}^* c_{30} + c_{20} c_{30}^*) \kappa_{13}^2\right) \kappa_{23}^3 \nonumber\\
    & \quad + \left(2 c_{30} \kappa_{13} (c_{10}^* \Delta - c_{30}^* \kappa_{13}) + c_{20} c_{20}^* (-2 \Delta^2 + \kappa_{13}^2)\right) \kappa_{23}^4 \nonumber\\
    & \quad + 2 c_{10}^* c_{20} \Delta \kappa_{23}^5 \nonumber\\
    & \quad + c_{10} (\kappa_{13}^2 + \kappa_{23}^2) \Big[c_{10}^* (2 \Delta^2 + \kappa_{13}^2) (\kappa_{13}^2 + \kappa_{23}^2) \nonumber\\
    & \quad + \Delta (c_{30}^* \kappa_{13} + c_{20}^* \kappa_{23}) (\kappa_{13}^2 + 2 \kappa_{23}^2)\Big] \nonumber\\
    & \quad + 2 \kappa_{13} \kappa_{23} \Omega \cos\left(\frac{\Delta t}{2}\right) \Big[ \left(- (c_{20}^* c_{30} + c_{20} c_{30}^*) \kappa_{13}^2 + 2 (-c_{20} c_{20}^* + c_{30} c_{30}^*) \kappa_{13} \kappa_{23}\right. \nonumber\\
    & \quad \left. + (c_{20}^* c_{30} + c_{20} c_{30}^*) \kappa_{23}^2\right) \Omega \cos\left(\frac{\Omega t}{2}\right) \nonumber\\
    & \quad + i \left(-c_{20}^* c_{30} \Delta + c_{20} c_{30}^* \Delta - c_{10}^* c_{20} \kappa_{13} + c_{10} c_{20}^* \kappa_{13} + c_{10}^* c_{30} \kappa_{23} - c_{10} c_{30}^* \kappa_{23}\right) \nonumber\\
    & \quad \times (\kappa_{13}^2 + \kappa_{23}^2) \sin\left(\frac{\Omega t}{2}\right)\Big] \nonumber\\
    & \quad + 2 \kappa_{13} \kappa_{23} \Omega \sin\left(\frac{\Delta t}{2}\right) \Big[ i (c_{20}^* c_{30} - c_{20} c_{30}^*) (\kappa_{13}^2 + \kappa_{23}^2) \Omega \cos\left(\frac{\Omega t}{2}\right) \nonumber\\
    & \quad + \Big(\kappa_{13}^2 \left(-c_{20}^* c_{30} \Delta - c_{20} c_{30}^* \Delta + c_{10}^* c_{20} \kappa_{13} + c_{10} c_{20}^* \kappa_{13}\right) \nonumber\\
    & \quad - \kappa_{13} \left(2 c_{20} c_{20}^* \Delta - 2 c_{30} c_{30}^* \Delta + c_{10}^* c_{30} \kappa_{13} + c_{10} c_{30}^* \kappa_{13}\right) \kappa_{23} \nonumber\\
    & \quad + \left(c_{20}^* c_{30} \Delta + c_{20} c_{30}^* \Delta + c_{10}^* c_{20} \kappa_{13} + c_{10} c_{20}^* \kappa_{13}\right) \kappa_{23}^2 - (c_{10}^* c_{30} + c_{10} c_{30}^*) \kappa_{23}^3\Big] \Big) \nonumber\\
    & \quad + (\kappa_{13}^2 + \kappa_{23}^2) (\kappa_{13}^2 + 2 \kappa_{23}^2) \Big[ - \Big((c_{10} \Delta + c_{30} \kappa_{13} + c_{20} \kappa_{23}) (c_{30}^* \kappa_{13} + c_{20}^* \kappa_{23}) \nonumber\\
    & \quad + c_{10}^* (c_{30} \Delta \kappa_{13} + c_{20} \Delta \kappa_{23} - c_{10} (\kappa_{13}^2 + \kappa_{23}^2))\cos\Omega t\Big) \nonumber\\
    & \quad - i (c_{10}^* (c_{30} \kappa_{13} + c_{20} \kappa_{23}) - c_{10} (c_{30}^* \kappa_{13} + c_{20}^* \kappa_{23})) \Omega \sin\Omega t\Big]\Big\} ,
\end{align}
%%%%%%%%%%%%%%%%%%%%%%%%%%%%%%%%%%%%%%%%%%%%%%%%%%%%%%%%%%%%%%%%%%%%%%%%%%%%
\begin{align}
n_4^\Lambda(t)&= \frac{e^{-\frac{1}{2} i \Delta t}}{(\kappa_{13}^2 + \kappa_{23}^2) \Omega^2} \nonumber\\ 
    &\quad  \times\Big\{ 
    \kappa_{23} \left( -c_{10}^* e^{i \Delta t} (\kappa_{20} \kappa_{13} - \kappa_{30} \kappa_{23}) + 
    c_{10} (-c_{20}^* \kappa_{13} + c_{30s} \kappa_{23}) \right) \Omega^2 \cos\left(\frac{\Omega t}{2}\right) \nonumber\\
    & \quad + i e^{i \Delta t} \kappa_{23} \left( c_{10}^* \Delta + c_{30s} \kappa_{13} + 
    c_{20}^* \kappa_{23} \right) (-\kappa_{20} \kappa_{13} + \kappa_{30} \kappa_{23}) \Omega 
    \sin\left(\frac{\Omega t}{2}\right) \nonumber\\
    & \quad + i \kappa_{23} \left( c_{10} \Delta + c_{30} \kappa_{13} + c_{20} \kappa_{23} \right) 
    (c_{20}^* \kappa_{13} - c_{30s} \kappa_{23}) \Omega \sin\left(\frac{\Omega t}{2}\right) \nonumber\\
    & \quad + e^{\frac{i \Delta t}{2}} \kappa_{13} \Big[ 
    c_{10} \left( c_{10}^* \Delta + c_{30s} \kappa_{13} + c_{20}^* \kappa_{23} \right) 
    (\kappa_{13}^2 + \kappa_{23}^2) \nonumber\\
    & \quad + \left( c_{30} \kappa_{13} + c_{20} \kappa_{23} \right) \left( -c_{30s} \Delta \kappa_{13} 
    - c_{20}^* \Delta \kappa_{23} + c_{10}^* (\kappa_{13}^2 + \kappa_{23}^2) \right) \nonumber\\
    & \quad + \Delta \Big[ \left( c_{10} \Delta + c_{30} \kappa_{13} + c_{20} \kappa_{23} \right) 
    (c_{30s} \kappa_{13} + c_{20}^* \kappa_{23}) \nonumber\\
    & \quad + c_{10}^* \left( c_{30} \Delta \kappa_{13} + c_{20} \Delta \kappa_{23} - 
    c_{10} (\kappa_{13}^2 + \kappa_{23}^2) \right) \Big] \cos\Omega t \nonumber\\
    & \quad + i \Delta \left( c_{10}^* (c_{30} \kappa_{13} + c_{20} \kappa_{23}) - 
    c_{10} (c_{30s} \kappa_{13} + c_{20}^* \kappa_{23}) \right) \Omega \sin\Omega t \Big\} ,
\end{align} 

\begin{align}
n_5^\Lambda(t)&=  \frac{e^{-\frac{i \Delta t}{2}}}{(\kappa_{13}^2 + \kappa_{23}^2) \Omega} \nonumber\\
&\quad \times\Big\{ 
    i \kappa_{23} \left( - c_{10}^* e^{i \Delta t} (c_{20} \kappa_{13} - c_{30} \kappa_{23}) 
    + c_{10} (c_{20}^* \kappa_{13} - c_{30}^* \kappa_{23}) \right) \Omega \cos \left( \frac{\Omega t}{2} \right) \nonumber\\
&\quad + e^{i \Delta t} \kappa_{23} \left( c_{10}^* \Delta + c_{30}^* \kappa_{13} + c_{20}^* \kappa_{23} \right) 
    \left( c_{20} \kappa_{13} - c_{30} \kappa_{23} \right) \sin \left( \frac{\Omega t}{2} \right) \nonumber\\
&\quad + \kappa_{23} \left( c_{10} \Delta + c_{30} \kappa_{13} + c_{20} \kappa_{23} \right) 
    \left( c_{20}^* \kappa_{13} - c_{30}^* \kappa_{23} \right) \sin \left( \frac{\Omega t}{2} \right) \nonumber\\
&\quad + e^{\frac{i \Delta t}{2}} \kappa_{13} \Big( 
        i \left( c_{10}^* (c_{30} \kappa_{13} + c_{20} \kappa_{23}) - c_{10} (c_{30}^* \kappa_{13} + c_{20}^* \kappa_{23}) \right) \Omega \cos\Omega t \nonumber\\
&\quad  - \Big( (c_{10} \Delta + c_{30} \kappa_{13} + c_{20} \kappa_{23}) (c_{30}^* \kappa_{13} + c_{20}^* \kappa_{23}) \nonumber\\
&\quad + c_{10}^* \left( c_{30} \Delta \kappa_{13} + c_{20} \Delta \kappa_{23} - c_{10} (\kappa_{13}^2 + \kappa_{23}^2) \right) \Big) \sin\Omega t
    \Big)
\Big\} ,
\end{align}

\begin{align}
n_6^\Lambda(t)&= \frac{1}{(\kappa_{13}^2 + \kappa_{23}^2)^2 \Omega^2} \Big\{ 
    \kappa_{13} \kappa_{23} \Big( 
        \kappa_{13}^2 \Big( 2 c_{30} c_{30}^* \Delta^2 
        - (c_{10}^* c_{30} + c_{10} c_{30}^*) \Delta \kappa_{13} \nonumber\\
        &\quad + (c_{10} c_{10}^* + c_{30} c_{30}^*) \kappa_{13}^2 
        - 2 c_{20} c_{20}^* (\Delta^2 + \kappa_{13}^2) \Big) \nonumber\\
        &\quad + \kappa_{13} \Big( 4 (c_{20}^* c_{30} + c_{20} c_{30}^*) \Delta^2 
        - (c_{10}^* c_{20} + c_{10} c_{20}^*) \Delta \kappa_{13} \nonumber\\
        &\quad + 3 (c_{20}^* c_{30} + c_{20} c_{30}^*) \kappa_{13}^2 \Big) \kappa_{23} \nonumber\\
        &\quad - \Big( 2 (-c_{20} c_{20}^* + c_{30} c_{30}^*) \Delta^2 
        + (c_{10}^* c_{30} + c_{10} c_{30}^*) \Delta \kappa_{13} \nonumber\\
        &\quad + (-2 c_{10} c_{10}^* + c_{20} c_{20}^* + c_{30} c_{30}^*) \kappa_{13}^2 \Big) \kappa_{23}^2 \nonumber\\
        &\quad - \Big( c_{10}^* c_{20} \Delta + c_{10} c_{20}^* \Delta 
        - 3 (c_{20}^* c_{30} + c_{20} c_{30}^*) \kappa_{13} \Big) \kappa_{23}^3 \nonumber\\
        &\quad + (c_{10} c_{10}^* + c_{20} c_{20}^* - 2 c_{30} c_{30}^*) \kappa_{23}^4 \Big) \nonumber\\
    &\quad + (\kappa_{13} - \kappa_{23}) (\kappa_{13} + \kappa_{23}) \Omega \cos\left( \frac{\Delta t}{2} \right) \Big( 
        \Big( (c_{20}^* c_{30} + c_{20} c_{30}^*) \kappa_{13}^2 \nonumber\\
        &\quad + 2 (c_{20} c_{20}^* - c_{30} c_{30}^*) \kappa_{13} \kappa_{23} 
        - (c_{20}^* c_{30} + c_{20} c_{30}^*) \kappa_{23}^2 \Big) \Omega \cos\left( \frac{\Omega t}{2} \right) \nonumber\\
        &\quad + i (c_{20}^* c_{30} \Delta - c_{20} c_{30}^* \Delta 
        + c_{10}^* c_{20} \kappa_{13} - c_{10} c_{20}^* \kappa_{13} \nonumber\\
        &\quad - c_{10}^* c_{30} \kappa_{23} + c_{10} c_{30}^* \kappa_{23}) 
        (\kappa_{13}^2 + \kappa_{23}^2) \sin\left( \frac{\Omega t}{2} \right) \Big) \nonumber\\
    &\quad + (\kappa_{13} - \kappa_{23}) (\kappa_{13} + \kappa_{23}) \Omega \sin\left( \frac{\Delta t}{2} \right) \Big( 
        - i (c_{20}^* c_{30} - c_{20} c_{30}^*) (\kappa_{13}^2 + \kappa_{23}^2) \Omega \cos\left( \frac{\Omega t}{2} \right) \nonumber\\
        &\quad + \Big( \kappa_{13}^2 (c_{20}^* c_{30} \Delta + c_{20} c_{30}^* \Delta 
        - c_{10}^* c_{20} \kappa_{13} - c_{10} c_{20}^* \kappa_{13}) \nonumber\\
        &\quad + \kappa_{13} \Big( 2 c_{20} c_{20}^* \Delta 
        - 2 c_{30} c_{30}^* \Delta + c_{10}^* c_{30} \kappa_{13} 
        + c_{10} c_{30}^* \kappa_{13} \Big) \kappa_{23} \nonumber\\
        &\quad - (c_{20}^* c_{30} \Delta + c_{20} c_{30}^* \Delta 
        + c_{10}^* c_{20} \kappa_{13} + c_{10} c_{20}^* \kappa_{13}) \kappa_{23}^2 \nonumber\\
        &\quad + (c_{10}^* c_{30} + c_{10} c_{30}^*) \kappa_{23}^3 \Big) \sin\left( \frac{\Omega t}{2} \right) \Big) \nonumber\\
    &\quad + \kappa_{13} \kappa_{23} (\kappa_{13}^2 + \kappa_{23}^2) \Big( 
        \Big( (c_{10} \Delta + c_{30} \kappa_{13} + c_{20} \kappa_{23}) (c_{30}^* \kappa_{13} + c_{20}^* \kappa_{23}) \nonumber\\
        &\quad + c_{10}^* (c_{30} \Delta \kappa_{13} + c_{20} \Delta \kappa_{23} 
        - c_{10} (\kappa_{13}^2 + \kappa_{23}^2)) \Big) \cos\Omega t \nonumber\\
        &\quad + i \Big( c_{10}^* (c_{30} \kappa_{13} + c_{20} \kappa_{23}) 
        - c_{10} (c_{30}^* \kappa_{13} + c_{20}^* \kappa_{23}) \Big) \Omega \sin\Omega t \Big) \Big\}
\end{align}

\begin{align}
n_7^\Lambda(t)&= \frac{1}{(\kappa_{13}^2 + \kappa_{23}^2) \Omega^2} \Big\{ 
    \Omega \sin\left( \frac{\Delta t}{2} \right) \Big( 
        \Big( (c_{20}^* c_{30} + c_{20} c_{30}^*) \kappa_{13}^2 \nonumber\\
        &\quad + 2 (c_{20} c_{20}^* - c_{30} c_{30}^*) \kappa_{13} \kappa_{23} 
        - (c_{20}^* c_{30} + c_{20} c_{30}^*) \kappa_{23}^2 \Big) \Omega \cos\left( \frac{\Omega t}{2} \right) \nonumber\\
        &\quad + i \Big( c_{20}^* c_{30} \Delta - c_{20} c_{30}^* \Delta 
        + c_{10}^* c_{20} \kappa_{13} - c_{10} c_{20}^* \kappa_{13} \nonumber\\
        &\quad - c_{10}^* c_{30} \kappa_{23} + c_{10} c_{30}^* \kappa_{23} \Big) (\kappa_{13}^2 + \kappa_{23}^2) 
        \sin\left( \frac{\Omega t}{2} \right) \Big) \nonumber\\
    &\quad + \Omega \cos\left( \frac{\Delta t}{2} \right) \Big( 
        i \Big( c_{20}^* c_{30} - c_{20} c_{30}^* \Big) (\kappa_{13}^2 + \kappa_{23}^2) \Omega \cos\left( \frac{\Omega t}{2} \right) \nonumber\\
        &\quad + \Big( \kappa_{13}^2 \Big( - c_{20}^* c_{30} \Delta - c_{20} c_{30}^* \Delta 
        + c_{10}^* c_{20} \kappa_{13} + c_{10} c_{20}^* \kappa_{13} \Big) \nonumber\\
        &\quad - \kappa_{13} \Big( 2 c_{20} c_{20}^* \Delta - 2 c_{30} c_{30}^* \Delta 
        + c_{10}^* c_{30} \kappa_{13} + c_{10} c_{30}^* \kappa_{13} \Big) \kappa_{23} \nonumber\\
        &\quad + \Big( c_{20}^* c_{30} \Delta + c_{20} c_{30}^* \Delta 
        + c_{10}^* c_{20} \kappa_{13} + c_{10} c_{20}^* \kappa_{13} \Big) \kappa_{23}^2 \nonumber\\
        &\quad - (c_{10}^* c_{30} + c_{10} c_{30}^*) \kappa_{23}^3 \Big) 
        \sin\left( \frac{\Omega t}{2} \right) \Big) 
\Big\}
\end{align}

\begin{align}
n_8^\Lambda(t)&= \frac{1}{2 \sqrt{3} (\kappa_{13}^2 + \kappa_{23}^2)^2 \Omega^2} 
\Big\{ 
    \kappa_{13}^4 \Big( 
        2 (c_{10} c_{10}^* + c_{20} c_{20}^* - 2 c_{30} c_{30}^*) \Delta^2 \nonumber \\ 
        & + 3 (c_{10}^* c_{30} + c_{10} c_{30}^*) \Delta \kappa_{13} - (c_{10} c_{10}^* - 2 c_{20} c_{20}^* + c_{30} c_{30}^*) \kappa_{13}^2 
    \Big) \nonumber\\
    %\quad
    &- 3 \kappa_{13}^3 \Big( 
        2 (c_{20}^* c_{30} + c_{20} c_{30}^*) \Delta^2 
        - (c_{10}^* c_{20} + c_{10} c_{20}^*) \Delta \kappa_{13} \nonumber\\
    & + (c_{20}^* c_{30} + c_{20} c_{30}^*) \kappa_{13}^2 
    \Big) \kappa_{23} \nonumber\\
    & + \kappa_{13}^2 \Big( 
        4 (c_{10} c_{10}^* - 2 c_{20} c_{20}^* + c_{30} c_{30}^*) \Delta^2 
        + 3 (c_{10}^* c_{30} + c_{10} c_{30}^*) \Delta \kappa_{13} \nonumber\\
    &+ 3 (-c_{20} c_{20}^* + c_{30} c_{30}^*) \kappa_{13}^2 
    \Big) \kappa_{23}^2 \nonumber\\
    &+ 3 \kappa_{13} \Big( 
        2 (c_{20}^* c_{30} + c_{20} c_{30}^*) \Delta^2 
        + (c_{10}^* c_{20} + c_{10} c_{20}^*) \Delta \kappa_{13} \nonumber\\
    &+ (c_{20}^* c_{30} + c_{20} c_{30}^*) \kappa_{13}^2 
    \Big) \kappa_{23}^3 \nonumber\\
    & + \Big( 
        2 (c_{10} c_{10}^* + c_{20} c_{20}^* - 2 c_{30} c_{30}^*) \Delta^2 
        + 3 (c_{10} c_{10}^* - c_{20} c_{20}^*) \kappa_{13}^2 
    \Big) \kappa_{23}^4 \nonumber\\
    & + 6 (c_{20}^* c_{30} + c_{20} c_{30}^*) \kappa_{13} \kappa_{23}^5 
    + 2 (c_{10} c_{10}^* + c_{20} c_{20}^* - 2 c_{30} c_{30}^*) \kappa_{23}^6 \nonumber\\
    & + 3 \kappa_{13} \Big( 
        2 \kappa_{23} \Omega \cos\left(\frac{\Delta t}{2}\right) 
        \Big[ 
            \Big( (c_{20}^* c_{30} + c_{20} c_{30}^*) \kappa_{13}^2 
            + 2 (c_{20} c_{20}^* - c_{30} c_{30}^*) \kappa_{13} \kappa_{23} \nonumber\\
    & - (c_{20}^* c_{30} + c_{20} c_{30}^*) \kappa_{23}^2 
            \Big) \Omega \cos\left( \frac{\Omega t}{2} \right) \nonumber\\
    &+ i \Big( 
        c_{20}^* c_{30} \Delta - c_{20} c_{30}^* \Delta 
        + c_{10}^* c_{20} \kappa_{13} - c_{10} c_{20}^* \kappa_{13} \nonumber\\
    & - c_{10}^* c_{30} \kappa_{23} + c_{10} c_{30}^* \kappa_{23} 
    \Big) (\kappa_{13}^2 + \kappa_{23}^2) \sin\left( \frac{\Omega t}{2} \right) 
        \Big] \nonumber\\
    & + 2 \kappa_{23} \Omega \sin\left(\frac{\Delta t}{2}\right) 
        \Big[ 
            -i (c_{20}^* c_{30} - c_{20} c_{30}^*) (\kappa_{13}^2 + \kappa_{23}^2) 
            \Omega \cos\left( \frac{\Omega t}{2} \right) \nonumber\\
    &+ \Big( 
        \kappa_{13}^2 \Big( 
            c_{20}^* c_{30} \Delta + c_{20} c_{30}^* \Delta - c_{10}^* c_{20} \kappa_{13} 
            - c_{10} c_{20}^* \kappa_{13} 
        \Big) \nonumber\\
    &+ \kappa_{13} \Big( 
            2 c_{20} c_{20}^* \Delta - 2 c_{30} c_{30}^* \Delta 
            + c_{10}^* c_{30} \kappa_{13} + c_{10} c_{30}^* \kappa_{13} 
        \Big) \kappa_{23} \nonumber\\
    &- \Big( 
            c_{20}^* c_{30} \Delta + c_{20} c_{30}^* \Delta 
            + c_{10}^* c_{20} \kappa_{13} + c_{10} c_{20}^* \kappa_{13} 
        \Big) \kappa_{23}^2 \nonumber\\
    &+ (c_{10}^* c_{30} + c_{10} c_{30}^*) \kappa_{23}^3 
    \Big) \sin\left( \frac{\Omega t}{2} \right) 
        \Big] \nonumber\\
    &+ \kappa_{13} (\kappa_{13}^2 + \kappa_{23}^2) 
    \Big[ -\Big( 
            (c_{10} \Delta + c_{30} \kappa_{13} + c_{20} \kappa_{23}) 
            (c_{30}^* \kappa_{13} + c_{20}^* \kappa_{23}) \nonumber\\
    &+ c_{10}^* \Big( 
            c_{30} \Delta \kappa_{13} + c_{20} \Delta \kappa_{23} 
            - c_{10} (\kappa_{13}^2 + \kappa_{23}^2) 
        \Big) 
        \Big) \cos\Omega t \nonumber\\
    &- i \Big( 
            c_{10}^* (c_{30} \kappa_{13} + c_{20} \kappa_{23}) 
            - c_{10} (c_{30}^* \kappa_{13} + c_{20}^* \kappa_{23}) 
        \Big) \Omega \sin\Omega t 
    \Big] 
\Big\},
\end{align} 
where $\Omega=\sqrt{\Delta ^2+\kappa _{13}^2+\kappa _{23}^2}$ . The SU(3) Bloch vectors of the $V$ and $\Xi$ configurations can be obtained following same procedure. 

%%%%%%%%%%%%%%%%%%%%%%%%%%%%%%%%%%%%%%%%%%%%

%\newpage 

%\section{Bloch equation for $\Lambda$, $V$ and $\Xi$ configurations}\label{secB}
\section*{Appendix B: Bloch equation for \texorpdfstring{$\Lambda$, $V$, and $\Xi$}{Lambda, V, and Xi} configurations}\label{secB}
\par 
To obtain the time derivative of the SU(3) Bloch vectors $\dot{n}_i^A(t)$ ($A=\Lambda, V, \Xi$), we shall make use of the Bloch equation of all three configurations and in this appendix we present their  derivation: 

\subsection*{i) \texorpdfstring{$\Lambda$}{Lambda} configuration:}
\par 
For the $\Lambda$ configuration, we consider their time derivative of Eq.\eqref{eq30}, 
\begin{align}
\frac{d\mathbf{n}^{\Lambda}(t)}{dt} =Tr\bigg[\boldsymbol{\lambda}\frac{d\rho_{\text{TLS}}^\Lambda}{dt}\bigg].
\end{align}
Using the Liouville equation, namely, 
\begin{align}
\frac{d\rho_{\text{TLS}}^{\Lambda}(t)}{dt}=-i\big[\rho_{\text{TLS}}^{\Lambda}(t),H^{\Lambda}_{\text{R}}(0)\big],
\end{align}
and the Hamiltonian Eq.\eqref{eq16}, it is straightforward to obtain the desired Bloch equation of the $\Lambda$ configuration (We choose the equal detuning condition, namely, $\Delta_{13}^\Lambda=\Delta_{23}^\Lambda=\Delta^{\Lambda}$): 
\begin{subequations}
\begin{align}
    \frac{dn_1^{\Lambda}(t)}{dt}&= -\Delta^{\Lambda} n_2^{\Lambda}(t)-\frac{1}{2}\kappa_{13} n_7^{\Lambda}(t), \\
    \frac{dn_2^{\Lambda}(t)}{dt}&= \Delta^{\Lambda} n_1^{\Lambda}(t)+\kappa_{23} n_3^{\Lambda}(t)-\frac{1}{2}\kappa_{13} n_6^{\Lambda}(t),\\
    \frac{dn_3^{\Lambda}(t)}{dt}&= -\kappa_{23} n_2^{\Lambda}(t)-\frac{1}{2}\kappa_{13} n_5(t),\\
    \frac{dn_4^{\Lambda}(t)}{dt}&= -\Delta^{\Lambda} n_5(t)+\frac{1}{2}\kappa_{23} n_7^{\Lambda}(t),\\
    \frac{dn_5^{\Lambda}(t)}{dt}&= \frac{1}{2}\kappa_{13} n_3^{\Lambda}(t)+\Delta^{\Lambda} n_4(t)-\frac{1}{2}\kappa_{23} n_6^{\Lambda}(t)+\frac{1}{2}\sqrt{3}\kappa_{13}n_8^{\Lambda}(t),\\
    \frac{dn_6^{\Lambda}(t)}{dt}&= \frac{1}{2}\kappa_{13} n_2^{\Lambda}(t)+\frac{1}{2}\kappa_{23} n_5^{\Lambda}(t),\\
    \frac{dn_7^{\Lambda}(t)}{dt}&= \frac{1}{2}\kappa_{13} n_1^{\Lambda}(t)-\frac{1}{2}\kappa_{23} n_4^{\Lambda}(t),\\
    \frac{dn_8^{\Lambda}(t)}{dt}&= \frac{1}{2}\sqrt{3}\kappa_{13}n_5^{\Lambda}(t),
\end{align}    
\end{subequations}
%\par
The time derivative of the Bloch vectors $\dot{n}_i^{\Lambda}(t)$ can be obtained by substituting the Bloch vectors from Eqs.(A.1-A.8) into the right side of Eqs.(A.11a-A.11h). The procedure is similar for the other two configurations and for completeness, we have outlined the derivation of the Bloch vectors in the remaining sections of the Appendix. 

\subsection*{ii) \texorpdfstring{$V$}{V} configuration:}

The unitary operator for the $V$ configuration is given by,  
\begin{align}
    U_V(t)=\exp\bigg[-\frac{i}{3}\big((2\Delta_{13}-\Delta_{12})V_3t+(2\Delta_{12}-\Delta_{13})U_3t\big)\bigg],
    \label{eq36}
\end{align}
and the corresponding Hamiltonian is given by, 
\begin{align}
\hat{H}_\text{R}^V(0)&=-i\dot{U}^\dag_\Lambda(t)U_V(t)+U^\dag_V(t)\hat{H}^\text{V}(t)U_V(t) \nonumber \\
&=\begin{bmatrix}
\frac{1}{3}(2\Delta_{13}-\Delta_{12}) & 0 & \kappa_{13} \\
0 & \frac{1}{3}(2\Delta_{12}-\Delta_{13}) & \kappa_{12} \\
\kappa_{13} & \kappa_{12} & -\frac{1}{3}(\Delta_{13}+\Delta_{12}),
%\label{eq36}
\end{bmatrix}.
\end{align}
where, $\Delta_{12}=2\omega_{12}+\omega_{13}-\Omega_{12}$ and $\Delta_{13}=2\omega_{13}+\omega_{12}-\Omega_{13}$ for the $V$ configuration, respectively. 
%par 
Proceeding similar way we obtain the Bloch equation for the $V$ configuration,
\begin{subequations}
\begin{align}
    \frac{dn_1^V(t)}{dt}&= -\kappa_{12}n_5^V(t) - \kappa_{13}n_7^V(t),\\
    \frac{dn_2^V(t)}{dt}&= \kappa_{12}n_4^V(t) - \kappa_{13}n_6^V(t),\\
    \frac{dn_3^V(t)}{dt}&= -\kappa_{13}n_5^V(t) + \kappa_{12}n_7^V(t,)\\
    \frac{dn_4^V(t)}{dt}&= - \kappa_{12}n_2^V(t) + \Delta^V n_5^V(t),\\
    \frac{dn_5^V(t)}{dt}&= \kappa_{12}n_1^V(t) + \kappa_{13}n_3^V(t) -\Delta^V n_5^V(t)+\sqrt{3} \kappa_{13}n_8^V(t), \\
    \frac{dn_6^V(t)}{dt}&= \kappa_{13}n_2^V(t) + \Delta^V n_7^V(t),\\
    \frac{dn_7^V(t)}{dt}&= \kappa_{13}n_1^V(t) - \kappa_{12}n_3^V(t) -\Delta^V n_6^V(t)+\sqrt{3} \kappa_{12}n_8^V(t),\\
    \frac{dn_8^V(t)}{dt}&= -\sqrt{3}\kappa_{13}n_5^V(t) - \sqrt{3}\kappa_{12}n_7^V(t),
\end{align}  
\end{subequations}
%for the $V$ configuration. 

\subsection*{iii) \texorpdfstring{$\Xi$}{Lambda} configuration:}
\par 
Finally for the $\Xi$ configuration using the unitary operator,  
\begin{align}
U_\Xi(t)=\exp\bigg[-\frac{i}{3}\big((2\Delta_{12}+\Delta_{23})U_3t+(\Delta_{12}+2\Delta_{23})T_3t\big)\bigg],
    \label{eq37}
\end{align}
we obtain the Hamiltonian of the $\Xi$ configuration, 
\begin{align}
\hat{H}_\text{R}^\Xi(0)&=-i\dot{U}^\dag_\Xi(t)U_\Xi(t)+U^\dag_\Xi(t)H^{\Xi}(t)U_\Xi(t) \nonumber \\
&=\begin{bmatrix}
\frac{1}{3}(\Delta_{12}+2\Delta_{23}) & \kappa_{23} & 0 \\
\kappa_{23} & \frac{1}{3}(\Delta_{12}-\Delta_{23}) & \kappa_{12} \\
0 & \kappa_{12} & -\frac{1}{3}(2\Delta_{12}+\Delta_{23})
%\label{eq36}
\end{bmatrix}. 
\end{align}
where the detuning are given by  $\Delta_{12}=2\omega_{12}+\omega_{23}-\Omega_{21}$ and $\Delta_{23}=2\omega_{23}+\omega_{32}-\Omega_{32}$, respectively. The corresponding Bloch equations are given by, 
\begin{subequations}
\begin{align}
    \frac{dn_1^{\Xi}(t)}{dt} &=\Delta^\Xi n_2^\Xi(t)  -\frac{1}{2}\kappa n_5^\Xi(t), \\
    \frac{dn_2^{\Xi}(t)}{dt} &=  \Delta^\Xi n_1^\Xi(t)  2\kappa n_3^\Xi(t) + \frac{1}{2}\kappa n_4^\Xi(t) \\
    \frac{dn_3^{\Xi}(t)}{dt} &= -2\kappa n_2^\Xi(t) + \frac{1}{2}\kappa n_7^\Xi(t), \\
    \frac{dn_4^{\Xi}(t)}{dt} &= \frac{1}{2}\kappa n_2^\Xi(t) + \Delta^\Xi n_5^\Xi(t) + \frac{1}{2}\kappa n_7^\Xi(t), \\
    \frac{dn_5^{\Xi}(t)}{dt} &= \frac{1}{2}\kappa n_1^\Xi(t) - 2\Delta^\Xi n_4^\Xi(t) - \frac{1}{2}\kappa n_6^\Xi(t), \\
    \frac{dn_6^{\Xi}(t)}{dt} &= \frac{1}{2}\kappa n_5^\Xi(t) + \Delta^\Xi n_7^\Xi(t), \\
    \frac{dn_7^{\Xi}(t)}{dt} &= -\frac{1}{2}\kappa n_3^\Xi(t) - \frac{1}{2}\kappa n_4^\Xi(t) - \Delta^\Xi n_6^\Xi(t) + \frac{1}{2}\sqrt{3}\kappa n_8^\Xi(t) , \\
    \frac{dn_8^{\Xi}(t)}{dt} &= -\frac{1}{2}\sqrt{3}\kappa n_7^\Xi(t),
\end{align}
\end{subequations}
where we have chosen equal value of the coupling parameters, $\kappa_{12}=\kappa_{23}=\kappa$ for the equidistant $\Xi$ configuration (i.e., ladder configuration). 

%======================================
\newpage 

% bibliography
\bibliographystyle{unsrt}
\bibliography{bloch-bibliography}

\end{document}